\documentclass[a4paper,11pt]{article}

\usepackage{amsmath,amssymb,array,amsthm,amsxtra,overpic,bbm,bbold,bm,epsfig,multirow,longtable,booktabs}
\usepackage{url}
\usepackage {indentfirst} 
\usepackage[utf8]{inputenc}
\usepackage[margin=1in]{geometry}
\usepackage[T1]{fontenc}
\usepackage{booktabs,multirow}
\usepackage{bbold}
\usepackage{subcaption}
\usepackage{amsmath}
\usepackage{mathtools,amssymb,braket,dsfont}
\usepackage{graphicx}
\usepackage{multicol}
\usepackage{multirow}
\usepackage{float}
\usepackage[dvipsnames]{xcolor}
\usepackage[final]{hyperref}
\hypersetup{
	colorlinks=true,
	linkcolor=orange,
	citecolor=blue,
	filecolor=magenta,
	urlcolor=blue
}
\usepackage[numbers, sort&compress]{natbib}
\usepackage[utf8]{inputenc}
\usepackage[T1]{fontenc}
\usepackage{comment}

\begin{document}
\begin{titlepage}
\begin{center}

{\LARGE \bf \boldmath Fermion masses and mixing in SO(10) GUT with a universal two-zero texture} \\[1cm]

Gao-Xiang~Fang$^{a,b,c,}$\footnote{E-mail: \texttt{fanggaoxiang21@mails.ucas.ac.cn}}
and
Ye-Ling~Zhou$^{a,}$\footnote{E-mail: \texttt{zhouyeling@ucas.ac.cn}}
\\[-2mm]
\end{center}
\vspace*{0.50cm}
\centerline{$^{a}$\it School of Fundamental Physics and Mathematical Sciences,}
\centerline{\it Hangzhou Institute for Advanced
Study, UCAS, Hangzhou 310024, China}
\centerline{$^{b}$\it Institute of Theoretical Physics, Chinese Academy of Sciences, Beijing 100190, China}
\centerline{$^{c}$\it University of Chinese Academy of Sciences, Beijing 100049, China}

\vspace*{1cm}

\begin{abstract}
{\noindent
We apply a universal two-zero texture (UTZT) to all mass matrices for matters in their flavour space in SO(10) GUT framework. This texture can be realised by assigning different charge for each family in a $Z_6$ symmetry. By fixing charged fermion masses at their best-fit values, we fit the rest 9 precisely measured observables (three angles and one CP-violating phase in the quark mixing, three angles in the lepton mixing, and two neutrino mass-squared differences) with {seven} model parameters.
The model fits all data of fermion masses and mixing very well and the leptonic CP-violating phase is predicted in the range $(90^\circ, 230^\circ)$. The model further predicts the right-handed neutrino masses, with the lightest and heaviest of order $10^9$ and $10^{12}$ GeV, respectively. 
Gauge unification and proton decay have been checked with the assumption of a breaking chain with two intermediate symmetries above the electroweak scale. It indicates that $\alpha_{\rm GUT}$ ranges in (0.022,0.032) as long as the assumption of economical choice of Higgs contents, and $M_{\rm GUT}$ should be bigger than $4.5\times 10^{15}$ GeV to meet the Super-K bound.  We show effective mass $m_{ee}$ for neutrinoless double beta decay, which provides us with a possibility to test grand unification with neutrinoless double beta decay experiments.
}
\end{abstract}

\end{titlepage}

\section{Introduction}

The flavour puzzle is a long-standing unresolved problem in particle physics. We still do not know the large difference between lepton mixing and quark mixing and whether the so many independent mixing parameters could be correlated with one another in an underlying mechanism beyond the Standard Model (SM). 

The texture-zero approach \cite{Fritzsch:1977za, Weinberg:1977hb, Wilczek:1977uh}, which was first proposed to calculate the Cabibbo angle and reduce free parameters in the quark sector, provides an efficient way to address the flavour puzzle. This approach assumes some entries of the quark Yukawa matrices (or equivalently, mass matrices) vanishes, and connect quark masses with the CKM mixing parameters. In this way, the number of free parameters of quark flavours is efficiently reduced.  A competitive pattern is the so-called four texture zeros \cite{Du:1992iy, Fritzsch:1995nx, Xing:1996hi}, that both up- and down-type quark Yukawa matrices are Hermitian and their (1,1) and (1,3) entries are zeros, i.e.,
\begin{eqnarray} \label{eq:two_zeros} 
	M_{f} &\sim& \left( \begin{matrix} 0 & \times & 0 \\
		\times  & \times & \times \\ 
		0 & \times & \times \\
	\end{matrix} \right) \;
\end{eqnarray}
for $f=u,d$, where the $(3,1)$ entry is also zero due to the assumption of Hermitian Yukawa matrices. This is a natural extension of the original Fritzsch texture which includes a third zero in the (2,2) entry \cite{Fritzsch:1977vd, Fritzsch:1979zq, Fritzsch:1986sn}. This texture takes advantages in its analytical simplicity \cite{Fritzsch:2002ga,Xing:2003yj} and solving the strong CP problem \cite{Xing:2015sva}; see recently in \cite{Fritzsch:2021ipb}. 

Texture zeros have been applied to the lepton sector by first assuming the light neutrino mass matrix taking the form as in Eq.~\eqref{eq:two_zeros} in the charged lepton flavour basis \cite{Frampton:2002yf, Xing:2002ta, Xing:2002ap}. A universal two-zero texture (UTZT) in the lepton sector, i.e., both $M_l$ and $M_\nu$ have two zero entries in the same position was proposed in \cite{Xing:2003zd}. 
Lepton flavour model with UTZT in the seesaw framework was constructed in \cite{Zhou:2012ds} within the context of Abelian discrete symmetries \cite{Grimus:2004hf}. 

Both four texture zeros in the quark sector and UTZT in the lepton sector fit the relevant flavour data very well until now. Motivated by their phenomenological successes, we will propose a unified two-zero textures in both quark and lepton sectors, a completely universal two-zero textures for all fermion mass matrices. Namely, the subscript $f$ in Eq.~\eqref{eq:two_zeros} will span for all quarks and leptons. To achieve this, we will work in the GUT framework. The grand unified version of UTZT has another advantage. It helps to reduce the large dimensionality of the parameter space in the flavour space.
We assume the gauge symmetry to be $SO(10)$ \cite{Fritzsch:1974nn} and take one of its key features that all quarks and leptons including the right-handed neutrino are assigned in the same sixteen-dimensional chiral representation. As a consequence, strong corrections of masses and mixing of all quarks and leptons are predicted \cite{Dutta:2004zh, Dutta:2005ni}; also see \cite{Babu:2018tfi, Babu:2018qca, Fu:2022lrn, Fu:2023mdu} with recent precision data taken into account. 

The model will be constructed in $SO(10) \times Z_6$ with the Pati-Salam symmetry \cite{Pati:1974yy} as an intermediate symmetry after GUT breaking  and before its breaking to the SM gauge symmetry. The texture zeros are realised in $Z_6$ symmetry following the method developed in \cite{Zhou:2012ds}. Applying Abelian discrete symmetries to realising flavour texture zeros in GUT models have been considered in \cite{Ferreira:2015jpa,Ivanov:2015xss,Ohlsson:2021lro,Lindestam:2021dyk}. 

The rest of this paper is organized as follows. In section~\ref{sec:2}, we
construct UTZT in $SO(10)$ GUTs with a $Z_6$ flavour symmetry.  
Section~\ref{sec:3} discusses analytical propoerties of UTZT in GUTs and performs the numerical analysis of the fermion masses and mixing. 
Section~\ref{sec:4} is contributed to explore the parameter space of intermediate scales  allowed by gauge unification and proton decay constraints under a certain breaking chain from $SO(10)$ to SM, where different copies of Higgs fields introduced along with $Z_6$ will be considered. 
We conclude in section~\ref{sec:5}.

\section{Realisation of universal two-zero flavour textures}\label{sec:2}

We construct the universal two-zero textures (UTZT) for fermion masses with gauge symmetry and flavour symmetry assumed to be $SO(10)$ and $Z_6$, respectively. We concentrate on the breaking chain  
\begin{eqnarray} \label{eq:breaking_chain}
SO(10) &\underset{\bf 54}{\overset{M_{\rm GUT}}\longrightarrow}& G_{422}^{\rm C} = SU(4)_c \times SU(2)_L \times SU(2)_R \times Z_2^{\rm C} \nonumber\\
&\underset{\bf 45}{\overset{M_2}\longrightarrow}& G_{3221} = SU(3)_c \times SU(2)_L \times SU(2)_R \times U(1)_X \nonumber\\ 
&\underset{\overline{\bf 126}}{\overset{M_1}\longrightarrow}& G_{\rm SM} = SU(3)_c \times SU(2)_L \times U(1)_Y\,,
\end{eqnarray} 
where symbols above and below each arrow refer to the energy scale of the corresponding symmetry breaking and the Higgs multiplet responsible for the breaking, respectively, and $Z_2^{\rm C}$ is the parity symmetry between left  particles and right charge-conjugate particles. All particles arrangements in $SO(10) \times Z_6$ and corresponding roles are listed in Table~\ref{tab:particle_content}. In addition, a CP symmetry, which will be spontaneously broken later, is introduced above the GUT scale. 
Below we first review some general features of fermion masses in $SO(10)$ and then discuss the construction of UTZT by imposing a $Z_6$ flavour symmetry.

\begin{table}[t] 
	\centering
	\begin{tabular}{ m{2cm}<{\centering} m{4cm}<{\centering} m{2.5cm}<{\centering} m{5.5cm}<{\centering} }
		\hline \hline \\[-4mm]
		 & Particles in $SO(10)$ & Charges in $Z_6$ & Roles in the theory \\[-4mm]\\\hline \\[-4mm]
		Fermions & $({\bf 16}_F^1, {\bf 16}_F^2, {\bf 16}_F^3)$ & $\{ 0,2,1 \}$ & Contains SM fermions \& RH neutrinos \\[-4mm]\\\hline 
		\\[-4mm]	
		& $({\bf 10}_H^1, {\bf 10}_H^2, {\bf 10}_H^3)$ & $\{ 4,3,2 \}$ & Generates fermion masses \\[-4mm]\\\cline{2-4} \\[-4mm]
		Higgses & $({\bf 120}_H^1, {\bf 120}_H^2, {\bf 120}_H^3)$ & $\{ 4,3,2 \}$ & Generates fermion masses \\[-4mm] \\\cline{2-4} \\[-4mm]
		& $(\overline{\bf 126}_H^1, \overline{\bf 126}_H^2, \overline{\bf 126}_H^3)$ & $\{ 4,3,2 \}$ & Generates fermion masses \& triggers LR symmetry breaking
		\\[-4mm] \\\cline{2-4} \\[-4mm]
		& ${\bf 54}_H$ & 0 & Triggers GUT symmetry breaking \\[-4mm] \\\cline{2-4} \\[-4mm]
		& ${\bf 45}_H$ & 0 & Triggers PS symmetry breaking \\[-4mm] \\\hline
		\hline
	\end{tabular}
	\caption{Particle content in $SO(10) \times Z_6$.}
	\label{tab:particle_content}
\end{table}

\subsection{Fermion masses in $SO(10)$ GUT and CP}

In $SO(10)$ GUTs, all fermions, including quarks and leptons as well as right-handed neutrinos, are introduced to explain light neutrino masses and are unified in a single matter field multiplet  ${\bf 16}$ of $SO(10)$. The matter field multiplet follows the representation product decomposition ${\bf 16} \times {\bf 16} = {\bf 10}_{\rm S} + \overline{\bf 126}_{\rm S} + {\bf 120}_{\rm A}$, where subscripts S and A correspond to symmetric and anti-symmetric combinations. To generate $SO(10)$-invariant Yukawa interaction, three Higgs multiplets are included ${\bf 10}_H$, $\overline{\bf 126}_H$ and ${\bf 120}_H$. In general, Yukawa couplings to generate fermion masses  can be arranged as
\begin{eqnarray} \label{eq:Yukawa_couplings}
	-{\cal L}_{\rm Y} = (A)_{\alpha\beta} \; {\bf 16}^\alpha_F {\bf 16}^\beta_F {\bf 10}_H \; + \; 
	(B)_{\alpha\beta} \; {\bf 16}^\alpha_F {\bf 16}^\beta_F \overline{\bf 126}_H \; + \; 
	i\, (C)_{\alpha\beta} \; {\bf 16}^\alpha_F {\bf 16}^\beta_F {\bf 120}_H \; + {\rm h.c.}\,,
\end{eqnarray}
where $\alpha,\beta=1,2,3$ run for three copies of flavours. $A$, $B$ and $C$ are in general $3 \times 3$ coupling matrices with $A$ and $B$ symmetric and $C$ antisymmetric. The CP symmetry imposes them to be real, as will be explained below, otherwise complex. Here, we have not considered copies of the Higgs in the flavour space, which will be specified in the next subsection. We are working in the non-SUSY framework and the complex conjugate of ${\bf 10}_H$ and ${\bf 120}_H$ transforms also as ${\bf 10}$- and ${\bf 120}$-plets of $SO(10)$. In the case that both ${\bf 10}_H$ and ${\bf 120}_H$ are real, Eq.~\eqref{eq:Yukawa_couplings} gives the most general Yukawa couplings in $SO(10)$. 
If ${\bf 10}_H$ and ${\bf 120}_H$ are complex, there might be additional couplings 
\begin{eqnarray} \label{eq:Yukawa_couplings_add}
	(A')_{\alpha\beta} \; {\bf 16}^\alpha_F {\bf 16}^\beta_F {\bf 10}_H^* \; + \; 
	i\, (C')_{\alpha\beta} \; {\bf 16}^\alpha_F {\bf 16}^\beta_F {\bf 120}_H^* \; + {\rm h.c.}
\end{eqnarray}
appearing in the Lagrangian. In this case, we forbid them by imposing an additional Peccei-Quinn (PQ) $U(1)$ symmetry \cite{Peccei:1977hh} as described in \cite{Babu:1992ia,Joshipura:2011nn, Bajc:2005zf}. 
In each case, Eq.~\eqref{eq:Yukawa_couplings} (following the proof in Appendix~\ref{app:1}) lead to the Dirac Yukawa coupling matrices taking the following structure \cite{Dutta:2004zh,Dutta:2005ni, Altarelli:2010at, Dutta:2009ij}:
\begin{eqnarray} \label{eq:Yukawa}
	Y_u &=& H + r_2 F + i r_3 G \,, \nonumber\\
	Y_d &=& r_1 (H+ F+ i G) \,, \nonumber\\
	Y_\nu &=& H - 3 r_2 F + i c_\nu G  \,, \nonumber\\
	Y_e &=& r_1 (H - 3 F + i c_e G) \,, 
\end{eqnarray}
where $H$ and $F$ are symmetric and $G$ is antisymmetric matrices. $H$, $F$ and $G$ are identical to $A^*$, $B^*$ and $C^*$ up to overall coefficients, respectively if both ${\bf 10}_H$ and ${\bf 120}_H$ are real. These Yukawa coupling matrices are written in the left-right convention as we are working in the non-supersymmetric GUT framework. Dirac masses for quarks and leptons are obtained after Higgses gain VEVs. 
The light neutrino mass matrix $M_\nu$ is obtained from type-(I+II) seesaw mechanism 
\begin{eqnarray} \label{eq:nu_mass}
	M_\nu &=& -m_L \, F + m_R \, Y_\nu F^{-1} Y_\nu^T \,,
\end{eqnarray}
where $m_L$ and $m_R$ are free and small mass parameters induced by Higgs VEVs. Here, we have parameterised the RH neutrino mass matrix as
\begin{eqnarray} \label{eq:N_mass}
	M_R = \frac{v^2}{2 m_R} F \,.
\end{eqnarray}

It is convenient to write Yukawa coupling terms in the Pati-Salam notation. In $SU(4)_c \times SU(2)_L \times SU(2)_R \subset SO(10)$, the ${\bf 16}_F$ fermion multiplet of $SO(10)$ is decomposed into two multiplets of the Pati-Salam gauge symmetry, which are denoted as  $\psi_L$ and $\psi_R^{\rm C}$, respectively, and
\begin{eqnarray}
	&&(\psi_L)_a{}_ i \,\sim ({\bf 4}, {\bf 2}, {\bf 1}) \,, \nonumber\\
	&&(\psi_R^{\rm C})^a{}_j \sim (\overline{\bf 4}, {\bf 1}, {\bf 2})\,,
\end{eqnarray}
where ${\rm C}$ in the superscript of a fermion represents the charge conjugation. All Higgs multiplets  are decomposed into bi-doublets of $SU(2)_L\times SU(2)_R$. More explicitly,
\begin{eqnarray}
	&&{\bf 10}_H \;\; \supset ({\bf 1}, {\bf 2}, {\bf 2}) \equiv \phi_{ij}\,, \nonumber\\
	&&\overline{\bf 126}_H \supset ({\bf 15}, {\bf 2}, {\bf 2}) \equiv \tilde{\phi}^a_b{}_{ij}\,,\nonumber\\
	&&{\bf 120}_H \supset ({\bf 1}, {\bf 2}, {\bf 2})' + ({\bf 15}, {\bf 2}, {\bf 2})' \equiv \eta_{ij} + \tilde{\eta}^a_b{}_{ij}\,.
\end{eqnarray} 
In this formula and above, $a$ and $b$ run for entries of the $SU(4)_c$ fundamental representation, and $i$ and $j$ run for entries of $SU(2)_L$ and $SU(2)_R$ fundamental representations, respectively. 
In the Pati-Salam symmetry, the Yukawa coupling terms, e.g., $(A)_{\alpha\beta} \; {\bf 16}^\alpha_F {\bf 16}^\beta_F {\bf 10}_H + i (C)_{\alpha\beta} \; {\bf 16}^\alpha_F {\bf 16}^\beta_F {\bf 120}_H + {\rm h.c.}$, are reduced to
\begin{eqnarray}
	\; (\overline{\psi^\alpha_R})^a{}_j \, (\psi^\beta_L)_a{}_i\, \varepsilon_{ii'}  \varepsilon_{jj'} \left[ (A)_{\alpha\beta} \, \phi_{i'j'} +  i \,(C)_{\alpha\beta} \, \eta_{i'j'} \right] & \nonumber\\
	+ 
	(\overline{\psi^\beta_L})^a{}_i \, (\psi^\alpha_R)_a{}_j\, \varepsilon_{ii'}  \varepsilon_{jj'}  \left[ (A)^*_{\alpha\beta} \, \phi_{i'j'}^* -  i \,(C)^*_{\alpha\beta} \, \eta_{i'j'}^* \right] &,
\end{eqnarray}
where $\epsilon = i \sigma_2$ has been used for singlet contraction from two doublets of $SU(2)$. 
The $SO(10)$ gauge symmetry includes an internal matter parity symmetry $Z_2^{\rm C}$. This parity is also called D parity in the reference. Explicit rules of the parity transformation in SO(10) and Pati-Salam group theories are found in \cite{Aulakh:2002zr}, which will not repeat here. General speaking, this parity in $SU(2)_L \times SU(2)_R$ appears to be  the following transformation 
\begin{eqnarray}
	&(\psi_L)_a{}_i \leftrightarrow (\psi_R^{\rm C})^a{}_j \,, \nonumber\\
	&\phi_{ij} \leftrightarrow \phi_{ji} \,, \quad
	\tilde{\phi}^a_b{}_{ij} \leftrightarrow \tilde{\phi}^b_a{}_{ji} \,, \quad
	\eta_{ij} \leftrightarrow - \eta_{ji} \,, \quad
	\tilde{\eta}^a_b{}_{ij} \leftrightarrow - \tilde{\eta}^b_a{}_{ji} \,.
\end{eqnarray}

We impose a CP symmetry above the GUT scale. All coupling matrices $A$, $B$ and $C$ are forced to be real in this symmetry \cite{Grimus:2006rk}. The CP transformation in $SU(2)_L \times SU(2)_R$ appears as 
\begin{eqnarray}
	&(\psi_L)_a{}_i \leftrightarrow (\psi_L^{\rm C})^a{}_j \,, \nonumber\\
	&\phi_{ij} \leftrightarrow \phi_{ji}^* \,, \quad
	\tilde{\phi}^a_b{}_{ij} \leftrightarrow \tilde{\phi}^b_a{}^*_{ji} \,, \quad
	\eta_{ij} \leftrightarrow - \eta^*_{ji} \,, \quad
	\tilde{\eta}^a_b{}_{ij} \leftrightarrow - \tilde{\eta}^b_a{}^*_{ji} 
\end{eqnarray} 
in the Pati-Salam convention, where transformation of spatial coordinates is dismissed. The CP symmetry is spontaneously broken by the VEV of ${\bf 120}$ \cite{Grimus:2006rk}, i.e., VEVs of $\eta_{ij}$ and $\tilde{\eta}^a_b{}_{ij}$. 
The CP symmetry combined with $Z_2^{\rm C}$ form a Klein symmetry. The latter includes one more parity transformation
\begin{eqnarray}
	&(\psi_L)_a{}_i \leftrightarrow (\psi_R)_a{}_i \,, \nonumber\\
	&\phi_{ij} \leftrightarrow \phi^*_{ij} \,, \quad
	\tilde{\phi}^a_b{}_{ij} \leftrightarrow \tilde{\phi}^b_a{}^*_{ij} \,, \quad
	\eta_{ij} \leftrightarrow \eta^*_{ij} \,, \quad
	\tilde{\eta}^a_b{}_{ij} \leftrightarrow \tilde{\eta}^b_a{}^*_{ij} \,.
\end{eqnarray}
This parity gives the permutation between left- and right-handed fermions and keep the Yukawa couplings Hermitian \cite{Xing:2015sva}.

\subsection{Texture zeros in $SO(10) \times Z_6$}\label{sec:2.2}

We introduce a $Z_6$ discrete symmetry in the flavour sector and consider to achieve UTZT for all Yukawa matrices. Texture zeros in general can be realised in Abelian discrete flavour symmetries \cite{Grimus:2004hf}. We follow the charge alignments given in \cite{Zhou:2012ds} by assuming a $Z_6$ flavour symmetry.
In a straightforward extension, we extend each Higgs multiplet into three copies, e.g., ${\bf 10}^k_H$ for $k=1,2,3$, etc. The renormalisable Yukawa couplings in Eq.~\eqref{eq:Yukawa_couplings} are then extended into
\begin{eqnarray} \label{eq:Yukawa_couplings_2}
-{\cal L}_{\rm Y} =
	(A_k)_{\alpha\beta} \; {\bf 16}^\alpha_F {\bf 16}^\beta_F {\bf 10}^k_H \; + \; 
	(B_k)_{\alpha\beta} \; {\bf 16}^\alpha_F {\bf 16}^\beta_F \overline{\bf 126}^k_H \; + \; 
	i\, (C_k)_{\alpha\beta} \; {\bf 16}^\alpha_F {\bf 16}^\beta_F {\bf 120}^k_H
\end{eqnarray}
for $k=1,2,3$, where $A_k$, $B_k$ and $C_k$ are all $3 \times 3$ coupling matrices with $A_k$ and $B_k$ symmetric and $C_k$ antisymmetric. 
We consider to arrange $Z_6$ charges for matter fields and Higgs fields as 
\begin{eqnarray} \label{eq:ABC}
	{\bf 16}^\alpha_F \;\quad &\sim& \quad \{ 0,2,1 \} \,, \nonumber\\[2mm]
	\left.\begin{array}{c} {\bf 10}^k_H \\[1mm] \overline{\bf 126}^k_H \\[1mm] {\bf 120}^k_H 
	\end{array} \right\}
	&\sim& \quad \{ 4,3,2 \} \,,
\end{eqnarray}
for $\alpha,k =1,2,3$, respectively. To be invariant under $Z_6$, all non-vanishing couplings in these terms should takes zero charge (mod 6). We check each term in Eq.~\eqref{eq:Yukawa_couplings_2} to see if such a condition is satisfied. 
We summarise that coupling matrices take the following textures,
\begin{eqnarray}\label{eq:matrices}
	A_1, B_1 \sim \begin{pmatrix} 0 & \times & 0 \\ \times & 0 & 0 \\ 0 & 0 & \times \end{pmatrix} \,, &&
	C_1  \sim \begin{pmatrix} 0 & \times & 0 \\ \times & 0 & 0 \\ 0 & 0 & 0 \end{pmatrix} \,, \nonumber\\
	A_2, B_2 \sim \begin{pmatrix} 0 & 0 & 0 \\ 0 & 0 & \times \\ 0 & \times & 0 \end{pmatrix} \,, &&
	C_2 \sim \begin{pmatrix} 0 & 0 & 0 \\ 0 & 0 & \times \\ 0 & \times & 0 \end{pmatrix} \,, \nonumber\\ 
	A_3, B_3  \sim \begin{pmatrix} 0 & 0 & 0 \\ 0 & \times & 0 \\ 0 & 0 & 0 \end{pmatrix} \,, &&
	C_3 =0\,.
\end{eqnarray} 
Here, a cross represents a non-vanishing entry in the matrix, referring to a zero $Z_6$ charge in the relevant Yukawa coupling. 

We check if UTZT is satisfied in all fermion Yukawa/mass matrices. 
All Dirac Yukawa coupling matrices $Y_f$ (for $f = u,d,e,\nu$) are linear combinations of matrices in Eq.~\eqref{eq:matrices}. Thus, they take forms of two-zero flavour textures as in Eq.~\eqref{eq:two_zeros}. 
Note these matrices are Hermitian as the CP symmetry is imposed. 
In the neutrino sector, the Majorana mass matrix for RH neutrino $M_R$ is a linear combination of $B_1$, $B_2$ and $B_3$. It is a real and symmetric matrix with two-zero texture as in Eq.~\eqref{eq:UTZT}. Below, we write all Yukawa matrices explicitly as, 
\begin{eqnarray} \label{eq:UTZT}
	Y_f= \zeta_f \left(
	\begin{array}{ccc}
		0 & C_f & 0 \\
		C_f^* & \tilde{B}_f & B_f \\
		0 & B_f^* & A_f \\
	\end{array}
	\right) \,,
	\label{eq:zero}
\end{eqnarray}
where $f=u,d,\nu,e$, entries $A_f$ and $\tilde{B}_f$ are real, and $B_f$ and $C_f$ are complex. We keep $A_f >0$ by extracting a sign parameter $\zeta_f = \pm 1$ out. 

For the light neutrino mass matrix $M_\nu$, via Type-(I+II) seesew formula in Eq.~\eqref{eq:nu_mass}, one can prove that the light neutrino mass matrix inherit the textures in Eq.~\eqref{eq:UTZT} via the type-(I+II) seesaw formula \cite{Fritzsch:1999ee,Hu:2011ac}. However, $M_\nu$ is not Hermitian but complex and symmetric. 

Before ending this section, we discuss some other possibilities to realise UTZT with the use of discrete symmetries. We first prove that given the PQ $U(1)$ symmetry, three copies of Higgs are the minimal requirement. Indeed, given any $Z_n$ symmetry with fermion ${\bf 16}_F^\alpha$ and Higgs $\{{\bf 10}, \overline{\bf 126}, {\bf 120} \}^k_H$ charges arranged with $q^F_\alpha$ and $q^H_k$ (both $q^F_\alpha$ and $q^H_k$ are integers less than $n$), algebras to achieve the two-zero textures are given in the following.
For any Higgs, their charges $q^H_k$  must satisfies, 
\begin{align} \label{eq:cond_1}
2q^F_1 + q^H_k \neq 0 ~({\rm mod}~ n)\,,  \nonumber\\
q^F_1 + q^F_3 + q^H_k \neq 0 ~({\rm mod}~ n) \,;
\end{align}
and there must be some Higgses with charges $q^H_{k_1}$, $q^H_{k_2}$, ..., satisfying
\begin{align} \label{eq:cond_2}
q^F_1 +q^F_2 + q^H_{k_1} = 0 ~({\rm mod}~ n)\,,  \nonumber\\
q^F_2 + q^F_3 + q^H_{k_2} = 0 ~({\rm mod}~ n) \,, \nonumber\\
2q^F_2 + q^H_{k_3} = 0 ~({\rm mod}~ n) \,, \nonumber\\
2q^F_3 + q^H_{k_4} = 0 ~({\rm mod}~ n) \,.
\end{align}
In order to distinguish flavours, $q^F_1 \neq q^F_2  \neq  q^F_3 \neq q^F_1$ (mod $n$) and $n \geqslant 3$. Taking this property into the above equation, we arrive at 
\begin{align}
q^H_{k_1} \neq q^H_{k_2} \neq q^H_{k_3} \neq q^H_{k_1},~ q^H_{k_2} \neq q^H_{k_4}  \neq q^H_{k_3}  ~({\rm mod}~ n) \,.
\end{align} 
That means $q^H_{k_1}$, $q^H_{k_2}$, $q^H_{k_3}$, as well as $q^H_{k_2}$, $q^H_{k_3}$, $q^H_{k_4}$, should be distinguishable charges. In the minimal case, $q^H_{k_1} = q^H_{k_4}$, and as a consequence, we are left with three copies of Higgses. 
Although three copies are the minimal requirement for UTZT, there are plenty of choices for $Z_n$ and charged assignments to satisfy the algebras in Eqs.~\eqref{eq:cond_1} and \eqref{eq:cond_2}. Namely, $Z_6$ is not unique in achieving the texture alignment.
For example, $Z_5$, with matter fields and Higgs fields arranged as following
	\begin{align} \label{eq:Z_5}
		\mathbf{16}_{F}^{\alpha}\sim\left\{1,2,4\right\},\quad\left.
		\begin{array}
			{c}\mathbf{10}_{H}^{k} \\[2mm]
			\overline{\mathbf{126}}_{H}^{k} \\[2mm]
			\mathbf{120}_{H}^{k}
		\end{array}\right\}\sim\left\{2,4,1\right\}
	\end{align}
can realise UTZT. 
In the case without the PQ symmetry, the number of Higgs copies can be reduced to 2. This is because ${\bf 10}^*_H$ and ${\bf 120}^*_H$ join in the Yukawa couplings as in Eq.~\eqref{eq:Yukawa_couplings_add}. The three different charges $q^H_{k_1}$, $q^H_{k_2}$ and $q^H_{k_3}$ can be arranged in the following way: two of them are charges of two distinguishable Higgses, and the third charge  refers to the charge conjugation of one of these Higgses. Applying the argument to our $Z_6$ model in Eq.~\eqref{eq:ABC}, the third copy of Higgses are not necessary to be introduced as they can be replaced by the charged conjugation of the first copy. For the $Z_5$ model in Eq.~\eqref{eq:Z_5}, the third copy are also not necessary as they can be replaced by the charge conjugation of the second copy. Reducing the copy of Higgses results in additional restriction on the Yukawa correlation between quarks and leptons, which is worth studying elsewhere.

\section{Fermion masses and mixing}\label{sec:3}

In this section, we will discuss correlations of masses and mixing between quarks and leptons both analytically and numerically. Heavy neutrino masses as required to generate light neutrino data via the seesaw mechanism will be predicted.

\subsection{Analytical derivation}

An remarkable advantage of the two-zero texture is the analytical calculability which enable us to analytically diagonalise the Yukawa matrices in terms of eigenvalues and the largest entry of the Yukawa matrix \cite{Xing:2003yj}. 
As all fermions are embedded in the same $SO(10)$ multiplet, all fermion masses are correlated. It is a non-trivial task to match fermion masses and mixing with their experimental data. 
We re-present Eq.~\eqref{eq:Yukawa} in the following form 
\begin{eqnarray} \label{eq:Y_e}
	Y_e &=& -\frac{4\, r_1}{r_2-1} {\rm Re} Y_u + \frac{r_2+3}{r_2-1} {\rm Re} Y_d + i\, c_e {\rm Im} Y_d \,, \nonumber\\
	Y_\nu &=& - \frac{3r_2+1}{r_2-1} {\rm Re} Y_u + \frac{4r_2}{r_1(r_2-1)}{\rm Re} Y_d + i \frac{c_\nu}{r_1} {\rm Im} Y_d \,,
\end{eqnarray}
where $r_1\, {\rm Im} Y_u = r_3\, {\rm Im} Y_d$ is satisfied. Restricted by the $Z_6$ symmetry, all Dirac Yukawa coupling matrices  $Y_f$ are Hermitian with two texture zeros, i.e.,
\begin{eqnarray}
	Y_f= \zeta_f \left(
	\begin{array}{ccc}
		0 & C_f & 0 \\
		C_f^* & \tilde{B}_f & B_f \\
		0 & B_f^* & A_f \\
	\end{array}
	\right) \,,
	\label{eq:zero}
\end{eqnarray}
where $f=u,d,\nu,e$, entries $A_f$ and $\tilde{B}_f$ are real, $B_f$ and $C_f$ are complex, $A_f>0$ and $\zeta_f = \pm1$. 
The light neutrino mass matrix $M_\nu$ and heavy neutrino mass matrix $M_R$ are given in Eqs.~\eqref{eq:nu_mass} and \eqref{eq:N_mass} with 
\begin{eqnarray}\label{eq:F}
	F &=&\frac{{\rm Re} Y_u}{r_2-1}- \frac{{\rm Re} Y_d}{r_1(r_2-1)}  \,. 
\end{eqnarray}
The resulting $M_\nu$ is a complex symmetric $3\times 3$ matrix with two-zero textures and $M_R$ is real.

We parametrise the quark sector in the following way. In the Yukawa matrix $Y_u$, which is in general complex, a minus sign of $B_u$ or $C_u$  can be rotated away by performing phase rotation with a phase $\pi$. This transformation does not change the real property of $F$, $H$ and $G$. With a phase rotation $P_u=\rm diag \{ 1, e^{i \phi_u}, e^{i \phi_u^\prime} \}$, the Hermitian Yukawa matrix $Y_u$ can be transformed into a real and symmetric matrix 
\begin{eqnarray}
	\overline{Y}_u \equiv P_u Y_u P_u^* =  \left(
	\begin{array}{ccc}
		0 & |C_u| & 0 \\
		|C_u| & \tilde{B}_u & |B_u| \\
		0 & |B_u| & A_u \\
	\end{array}
	\right) \,.
\end{eqnarray} 
Without loss of generality, we can set $\zeta_u=+1$, and $\tilde{B}_u$ could be either positive and negative. They are correlated with up-type quark Yukawa couplings in the following way \cite{Fritzsch:2002ga,Fritzsch:2021ipb}
\begin{eqnarray} \label{eq:ABC_u}
	\tilde{B}_u &=& -\eta_u y_u + \eta_u y_c +y_t - A_u \,,\nonumber\\
	|B_u| &=& \sqrt{(A_u + \eta_u y_u)(A_u - \eta_u y_c) (y_t - A_u)/A_u} \,, \nonumber\\
	|C_u| &=& \sqrt{y_u y_c y_t / A_u} \,,
\end{eqnarray}
where $\eta_u = \pm 1$ is an undetermined sign. The real orthogonal matrix, which is used in the diagonalisation 
\begin{eqnarray}
	\overline{Y}_u = O_u \hat{Y}_u O_u^T 
\end{eqnarray}
with $\hat{Y}_u = {\rm diag}\{ -\eta_u y_u, \eta_u y_c, y_t\}$, can be explicitly expressed in terms of $A_u$ and $y_u$, $y_c$, $y_t$ as \cite{Fritzsch:2002ga,Fritzsch:2021ipb}
\begin{eqnarray}
	O_u= 
	\begin{pmatrix}
		\sqrt{\frac{y_c y_t (A_u+\eta_u  y_u)}{A_u (y_u+y_c) (\eta_u  y_u+y_t)}} & 
		\eta_u  \sqrt{\frac{y_u y_t (A_u-\eta_u  y_c)}{A_u (y_u+y_c) (y_t-\eta_u  y_c)}} & 
		\sqrt{\frac{y_u y_c (y_t-A_u)}{A_u (\eta_u  y_u+y_t) (y_t-\eta_u  y_c)}} \\[5mm]
		-\eta_u  \sqrt{\frac{y_u (A_u+\eta_u  y_u)}{(y_u+y_c) (\eta_u  y_u+y_t)}} & 
		\sqrt{\frac{y_c (A_u-\eta_u  y_c)}{(y_u+y_c) (y_t-\eta_u  y_c)}} & 
		\sqrt{\frac{y_t (y_t-A_u)}{(\eta_u  y_u+y_t) (y_t-\eta_u  y_c)}} \\[5mm]
		\eta_u  \sqrt{\frac{y_u (y_t-A_u) (A_u-\eta_u  y_c)}{A_u (y_u+y_c) (\eta_u  y_u+y_t)}} & 
		-\sqrt{\frac{y_c (y_t-A_u) (A_u+\eta_u  y_u)}{A_u (y_u+y_c) (y_t-\eta_u  y_c)}} & 
		\sqrt{\frac{y_t (A_u+\eta_u  y_u) (A_u-\eta_u  y_c)}{A_u (\eta_u  y_u+y_t) (y_t-\eta_u  y_c)}}
	\end{pmatrix} \,.
\end{eqnarray}
The down-type quark Yukawa matrix $Y_d$ is in general complex. With a phase rotation  $P_d={\rm diag}\{ 1, e^{i \phi_d}, e^{i \phi_d^\prime} \}$, it can be transformed into a real and symmetric matrix
\begin{eqnarray}
	\overline{Y}_d \equiv \zeta_d P_d Y_d P_d^* =  \left(
	\begin{array}{ccc}
		0 & |C_d| & 0 \\
		|C_d| & \tilde{B}_d & |B_d| \\
		0 & |B_d| & A_d \\
	\end{array}
	\right) \,.
\end{eqnarray} 
On the right-hand side, one can perform a similar transformation as for $Y_u$ with the orthogonal matrix $O_d$ given by 
\begin{eqnarray}
	O_d= 
	\begin{pmatrix}
		\sqrt{\frac{y_s y_b (A_d+\eta_d  y_d)}{A_d (y_d+y_s) (\eta_d  y_d+y_b)}} & 
		\eta_d  \sqrt{\frac{y_d y_b (A_d-\eta_d  y_s)}{A_d (y_d+y_s) (y_b-\eta_d  y_s)}} & 
		\sqrt{\frac{y_d y_s (y_b-A_d)}{A_d (\eta_d  y_d+y_b) (y_b-\eta_d  y_s)}} \\[5mm]
		-\eta_d  \sqrt{\frac{y_d (A_d+\eta_d  y_d)}{(y_d+y_s) (\eta_d  y_d+y_b)}} & 
		\sqrt{\frac{y_s (A_d-\eta_d  y_s)}{(y_d+y_s) (y_b-\eta_d  y_s)}} & 
		\sqrt{\frac{y_b (y_b-A_d)}{(\eta_d  y_d+y_b) (y_b-\eta_d  y_s)}} \\[5mm]
		\eta_d  \sqrt{\frac{y_d (y_b-A_d) (A_d-\eta_d  y_s)}{A_d (y_d+y_s) (\eta_d  y_d+y_b)}} & 
		-\sqrt{\frac{y_s (y_b-A_d) (A_d+\eta_d  y_d)}{A_d (y_d+y_s) (y_b-\eta_d  y_s)}} & 
		\sqrt{\frac{y_b (A_d+\eta_d  y_d) (A_d-\eta_d  y_s)}{A_d (\eta_d  y_d+y_b) (y_b-\eta_d  y_s)}}
	\end{pmatrix} \,.
\end{eqnarray}
Namely, 
\begin{eqnarray}
	Y_d = \zeta_d P_d^* O_d \hat{Y}_d O_d^T P_d \,,
\end{eqnarray}
where $\hat{Y}_d = {\rm diag}\{ -\eta_d y_d, \eta_d y_s, y_b\}$. 
The CKM mixing matrix is given by 
\begin{eqnarray}
	V_{\rm CKM} = O_u^T P_u P_d^* O_d \,.
\end{eqnarray}

It is useful to reparametrise $A_u$ and $A_d$ in the form 
$A_u = y_t (r + \epsilon)$ and $A_d = y_b (r - \epsilon)$. From the numerical analysis, we find that in order to satisfy experimental data, the restriction $|\epsilon| < 0.03$ is satisfied. From experimental
quark data, the following hierarchical relations exist among mixing parameters:
\begin{eqnarray} \label{eq:Cabibbo_angle}
	y_u : y_c : y_t &\sim& \theta_C^8 : \theta_C^4 : \theta_C^0 \,, \nonumber\\
	y_d : y_s : y_b &\sim& \theta_C^8 : \theta_C^6 : \theta_C^3 \,, \\
	\theta_{13}^q : \theta_{23}^q : \theta_{12}^q &\sim& \theta_C^3 : \theta_C^2 : \theta_C^1 \,, \nonumber
\end{eqnarray}
where $\theta_C$ is the Cabibbo angle. Using these relations, approximately, we derive the three mixing angles as 
\begin{eqnarray} \label{eq:approximation}
	\sin \theta _{12}^q &\approx& \left| \eta_d \sqrt{\frac{y_d}{y_s}}-\eta_u \sqrt{\frac{y_u}{y_c}} \left[ (1-r) e^{i \phi_2}+r e^{i \phi_1}\right] \right| \,, \nonumber\\
	\sin \theta _{13}^q &\approx& \left| \frac{\sqrt{y_d y_s}}{y_b} \sqrt{\frac{1-r}{r}}-\eta_u \sqrt{\frac{y_u}{y_c}} \left[\sqrt{(1-r) r} \left(e^{i \phi_1}\!-\!e^{i \phi_2}\right)+\frac{\epsilon}{2\sqrt{(1-r) r}} \left(e^{i \phi_1}\!+\!e^{i \phi_2}\right)\right] \right| \,,\nonumber\\ 
	\sin \theta _{23}^q &\approx& \left| \frac{\epsilon  \left(e^{i \phi_1}+e^{i \phi_2}\right)}{2 \sqrt{(1-r) r}}+\sqrt{(1-r) r} \left(e^{i \phi_1}-e^{i \phi_2}\right) \right| \,,
\end{eqnarray}
where $\phi_1=\phi_u-\phi_d,\, \phi_2=\phi_u^\prime-\phi_d^\prime$. As seen in Eq.~\eqref{eq:Y_e}, $Y_e$ can be written in terms of $Y_u$ and $Y_d$ with coefficients $r_1$, $r_2$ and $c_e$.
On the other hand, entries of $Y_e$ satisfy correlations similar to those in Eq.~\eqref{eq:ABC_u}, 
\begin{eqnarray} \label{eq:ABC_e}
	\tilde{B}_e &=& -\eta_e y_e + \eta_e y_\mu +y_\tau - A_e \,,\nonumber\\
	|B_e| &=& \sqrt{(A_e + \eta_e y_e)(A_e - \eta_e y_\mu) (y_\tau - A_e)/A_e} \,, \nonumber\\
	|C_e| &=& \sqrt{y_e y_\mu y_\tau / A_e} \,,
\end{eqnarray}
where $A_e$ is the absolute value of the $(3,3)$ entry of $Y_e$. 
This parametrisation helps us to simplify our numerical analysis in the next subsection. 

In the neutrino sector, however, $M_\nu$ is a complex and symmetric matrix, the phase rotation cannot transform $M_\nu$ to a real matrix. The unitary matrix to diagonalise $M_\nu$ has to be performed numerically. For RH neutrinos, the Majorana mass matrix $M_R$, which is proportional to $F$ as a linear combination of ${\rm Re}\, Y_u$ and ${\rm Re}\, Y_d$, is a real matrix with two texture zeros. It can be analytically diagonalised by a real orthogonal matrix $O_R$, which takes a similar form as $O_{u,d,e}$ but replacing the yukawa couplings by relevant RH neutrino masses $M_{N_1}$, $M_{N_2}$ and $M_{N_3}$. We will not repeat here.

%%%%%%%%%%%%%%%%%%%%%%%%%%%%%%%%%%%%%%%%%%%%%%%%%%%%%%%%%%%%%%%%%%%%%%%%%%%%%%%%%%%%%%%%%%%%%%%%%%%%%%%%%%%%%%%%%%%%%%%%%%%%%%%%%%%%%%%%%%%%%%%%%%%%%%%%%%%%%%%%%%%%%%%%%%%%%%%%%

\subsection{Numerical analysis}\label{sec:3_2}

We describe our numerical analysis in this subsection.
To simplify the analysis, we fix the numerical best-fit (bf) results of charged fermion Yukawa couplings, 
\begin{eqnarray}
	&&
	y_u^{\rm bf} = 2.54 \times 10^{-6}, \quad\;
	y_c^{\rm bf} = 1.37 \times 10^{-3}, \quad\;
	y_t^{\rm bf} =  0.428, \nonumber\\
	&&
	y_d^{\rm bf} = 6.56 \times10^{-6}, \quad\;
	y_s^{\rm bf} = 1.24 \times10^{-4}, \quad\;
	y_b^{\rm bf} =5.7 \times 10^{-3}, \nonumber\\
	&&
	y_e^{\rm bf} = 2.70341 \times 10^{-6}, \quad\;
	y_\mu^{\rm bf} = 5.70705 \times 10^{-4}, \quad\;
	y_\tau^{\rm bf} = 9.702 \times 10^{-3}. 
\end{eqnarray}
For these values, we do not consider the errors of them, which means pulls for all charged fermion Yukawa couplings are fixed at zero explicitly. 
Best-fit and $1\sigma$ values of three mixing angles and one CP-violating phase in the CKM mixing matrix are taken to be
\begin{eqnarray}
&	\theta^{q}_{12} = 13.028^\circ \pm 0.034^\circ \,,\;
	\theta^{q}_{23} = 2.783^\circ \pm 0.034^\circ \,,\;
	\theta^{q}_{13} = 0.241^\circ \pm 0.007^\circ \,, \nonumber\\
&	\delta^{q}=69.52^\circ \pm 3.09^\circ \,.
\end{eqnarray}
These values were calculated at the GUT scale ($2\times 10^{16}$~GeV) from a non-SUSY
scenario as done in \cite{Babu:2016bmy}.

In the quark sector, because the condition $r_1\, {\rm Im} Y_u = r_3\, {\rm Im} Y_d$ is always satisfied, which means $\frac{\rm Im (Y_u)_{12}}{\rm Im (Y_d)_{12}}=\frac{\rm Im (Y_u)_{23}}{\rm Im (Y_d)_{23}}=\frac{r_3}{r_1}$, so the value of $\phi_d^\prime$ is fully determined by $\{\phi_u, \phi_u^\prime, \phi_d\}$.
Once Yukawa couplings $y_f$ for $f=u,c,t,d,s,b$ are fixed at their best-fit values, $Y_u$ and $Y_d$ are fully determined by the free parameters {$\{A_u, A_d, \phi_1, \phi_2 \}$} up to sign parameters $\{\eta_u, \eta_d, \zeta_u, \zeta_d\}$, where $\zeta_u =1$ is fixed without loss of generality.  
$\zeta_d$ does not influence  masses and mixing in the quark sector but contributes to the lepton sector via Eq.~\eqref{eq:Y_e}. We scan 
\begin{eqnarray} \label{eq:scan_quark}
	A_u/y_t, A_d/y_b \in (0, 1) \,, \quad
	\phi_1, \phi_2,  \in (0, 2\pi) \,.
\end{eqnarray}

After $Y_u$ and $Y_d$ are determined by the above fitting procedure in the quark sector, all non-vanishing entries of $Y_e$, i.e., $A_e$, $B_e$, $\tilde{B}_e$ and $C_e$, are functions of $r_1$, $r_2$ and $c_e$. We fix $y_e$, $y_\mu$ and $y_\tau$ at their best-fit values, $r_1$, $r_2$ and $c_e$ are determined with the help of Eq.~\eqref{eq:ABC_e} up to a sign difference $\eta_e$. Then $Y_e$ is fixed, and we can determine the unitary matrix $U_e$ in the diagonalisation $U_e^\dagger Y_e U_e = \text{diag}\{y_e,y_\mu,y_\tau\}$.

In the neutrino sector, as seen in Eq.~\eqref{eq:Y_e}, the only undetermined parameter in $Y_\nu$ is $c_\nu$. In Eq.~\eqref{eq:nu_mass}, the light neutrino mass matrix, $M_\nu$, is determined by two further parameters $m_L$ and $m_R$. We scan these parameters in the following regions 
\begin{eqnarray} \label{eq:scan_lepton}
c_\nu \in (10^{-2}, 10)\,,\quad 
m_L, m_R \in (10^{-1}, 10^{2})~{\rm eV} 
\end{eqnarray}
in the logarithmic scale to obtain $M_\nu$. Then the diagonalisation $V_\nu^\dagger M_\nu V_\nu^*=\text{diag}\{m_1,m_2,m_3\}$ provides the neutrino mass eigenvalues and unitary matrix $V_\nu$. Finally, the PMNS matrix is given by $U_{\rm PMNS}=V_e^\dagger V_\nu$ and the three lepton mixing angles are given by 
\begin{eqnarray}
	\sin \theta_{13}^l=|(U_{\rm PMNS})_{e3}|, \quad \tan \theta_{12}^l=\bigg|\frac{(U_{\rm PMNS})_{e2}}{(U_{\rm PMNS})_{e1}} \bigg|, \quad \tan \theta_{23}^l=\bigg|\frac{(U_{\rm PMNS})_{\mu 3}}{(U_{\rm PMNS})_{\tau 3}} \bigg|.
\end{eqnarray}
On the experimental side, we take their global bf values (without including SK atmospheric data) from NuFIT 5.3 \cite{Esteban:2020cvm, nufit5.3}\footnote{The most recent version, NuFIT 6.0 \cite{nufit6.0}, was released during our work. The main difference is that the best-fit value of the CP-violating phase shifting from $197^\circ$ to $177^\circ$, as well as a smaller deviation of $\theta_{23}$ from the maximal mixing value, i.e., from $49.1^\circ$ to $48.5^\circ$. These differences have little influence to our scan and we keep our fitting in v5.3.} and average the positive and negative $1\sigma$ errors, as well as values of two mass-squared differences $\Delta m^2_{21}=m_2^2-m_1^2$ and $\Delta m^2_{31}=m_3^2-m_1^2$
\begin{eqnarray} \label{eq:mass square}
	&\Delta m^2_{21} = (7.41 \pm 0.21) \times 10^{-5} {\rm eV}^2\,, \;
	\Delta m^2_{31} = (2.511 \pm 0.027) \times 10^{-3} {\rm eV}^2 \,, \nonumber\\
	&\theta_{12}^l = 33.66^\circ \pm 0.73^\circ\,, \;
	\theta_{23}^l = 49.1^\circ \pm 1.3^\circ\,, \;
	\theta_{13}^l = 8.54^\circ \pm 0.11^\circ\,, \;
\end{eqnarray}
for the normal mass ordering (NO, i.e., $m_1 < m_2 < m_3$) into account. The up-to-date experimental constraint on the Dirac CP-violating phase $\delta^l$, $\delta^l= 197^\circ \pm 41^\circ$, which is still weak, is not included in the fit. Instead, we treat $\delta^l$ as a prediction of the model.
 We will not discuss inverted ordering (i.e., $m_3 < m_1 < m_2$), because a preliminary scan suggests that our model does not favour this configuration well. Furthermore, we do not consider the small flavour-dependent RG running effect due to the suppression of charged lepton Yukawa coupling.

We count for numbers of free parameters introduced in the model and independent observables to fit the data. Once the charged fermion masses are fixed, we are left with five free parameters $\{A_u,A_d,\phi_u,\phi_u^\prime,\phi_d\}$ for scan and two signs $\{\eta_u,\eta_d\}$ in the quark sector, and there are four observables $\{\theta_{12}^q,\theta_{13}^q,\theta_{23}^q,\delta^q\}$ to fit. In the neutrino sector, we scan another three free parameters $\{c_\nu, m_L, m_R\}$ and two signs $\{\eta_e,\zeta_d\}$, and there are five observables $\{\theta_{12}^l,\theta_{13}^l,\theta_{23}^l,\Delta m^2_{21},\Delta m^2_{31}\}$ to fit.
In summary, there are {seven} free parameters 
\begin{eqnarray} \label{eq:parameters}
{\rm para}_m \in \{A_u,A_d,\phi_1,\phi_2,c_\nu,m_L, m_R\} \,.
\end{eqnarray}
These parameters are used to fit nine independent observables 
\begin{eqnarray} \label{eq:predictions}
{\rm obs}_n \in 
\{\theta_{12}^q,\theta_{13}^q,\theta_{23}^q,\delta^q,\theta_{12}^l,\theta_{13}^l,\theta_{23}^l,\Delta m^2_{21}, \Delta m^2_{31}\}
\,.
\end{eqnarray} 
Other observables are treated as predictions of the model. 

\begin{figure}[t] 
	\begin{center} 
		\includegraphics[width=0.45\textwidth]{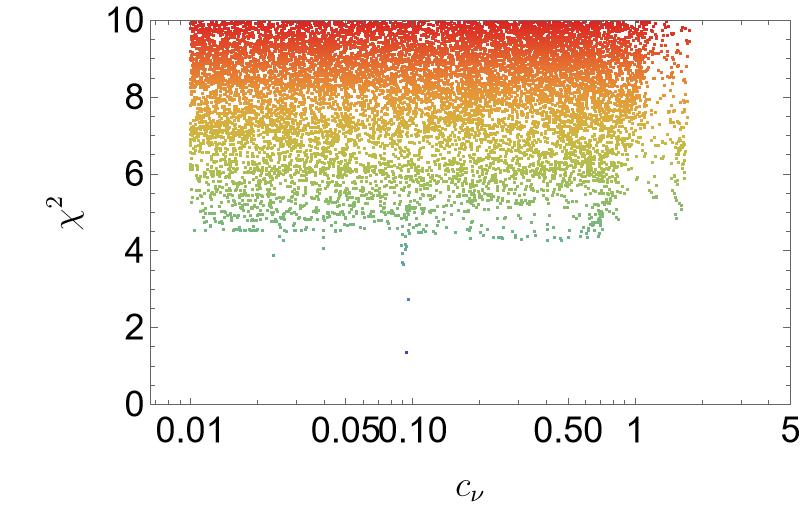}
		\includegraphics[width=0.45\textwidth]{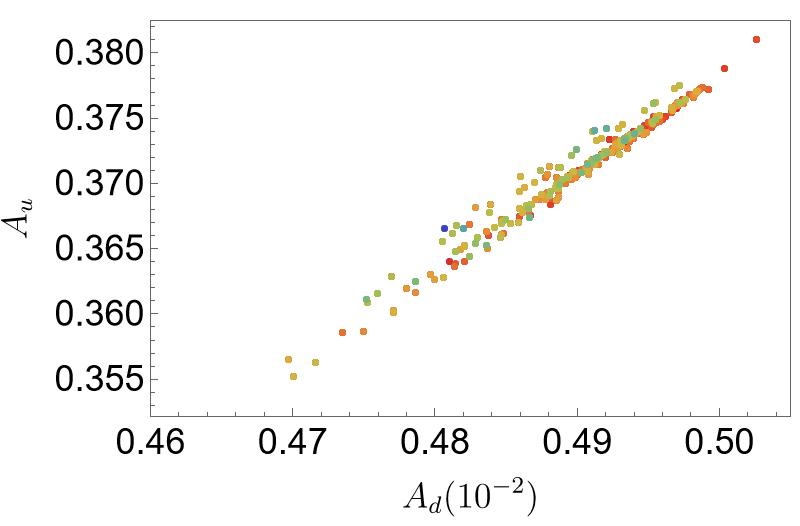}
		\includegraphics[width=0.45\textwidth]{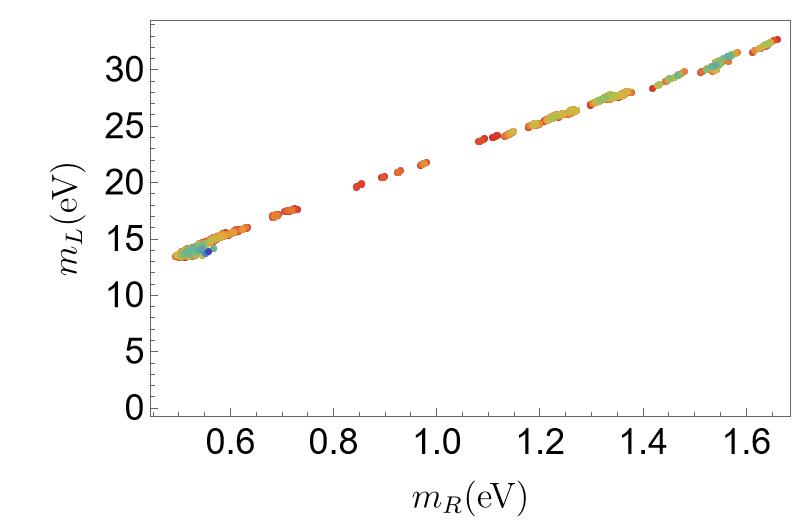}
		\includegraphics[width=0.45\textwidth]{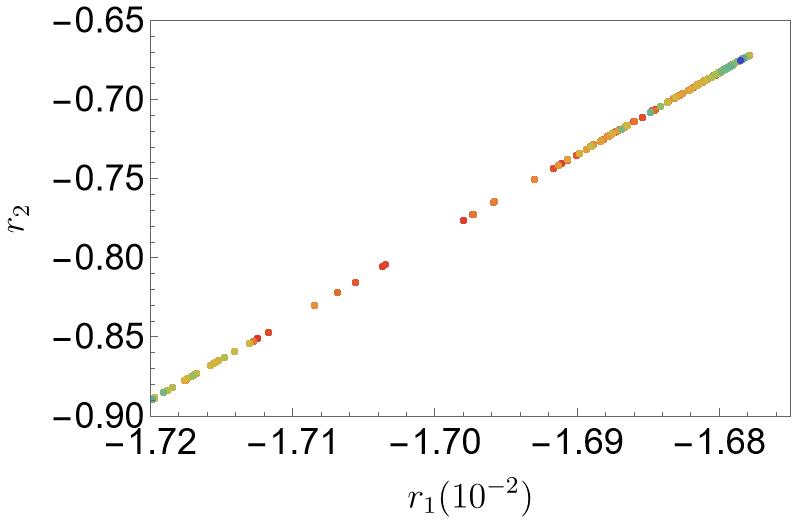}		
		\includegraphics[width=0.4\textwidth]{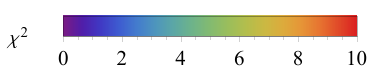}
		\caption{Model parameters in the model with $\chi^2 \leq 10$.}
		\label{fig:parameters}
	\end{center}
\end{figure}

\begin{figure}[t] 
	\begin{center} 
		\includegraphics[width=0.45\textwidth]{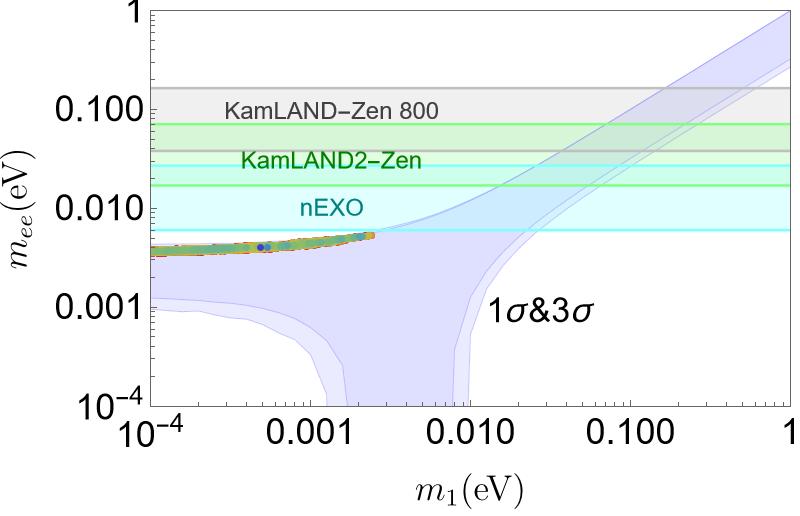}
		\includegraphics[width=0.45\textwidth]{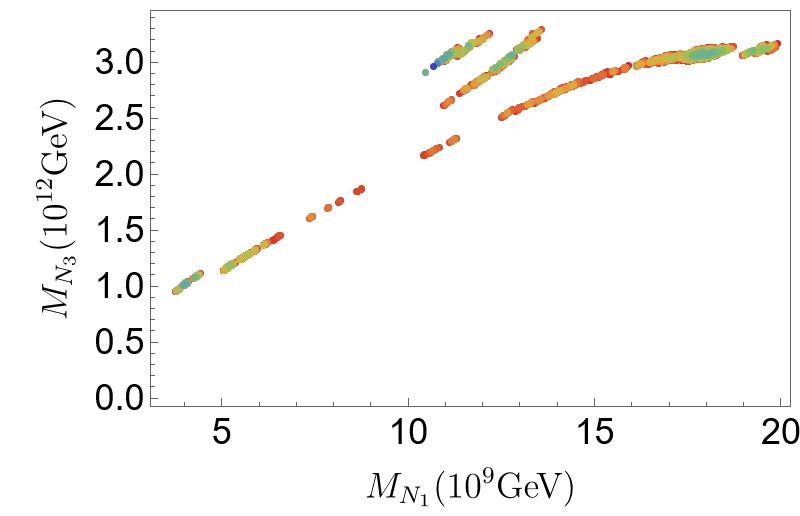}
		\includegraphics[width=0.45\textwidth]{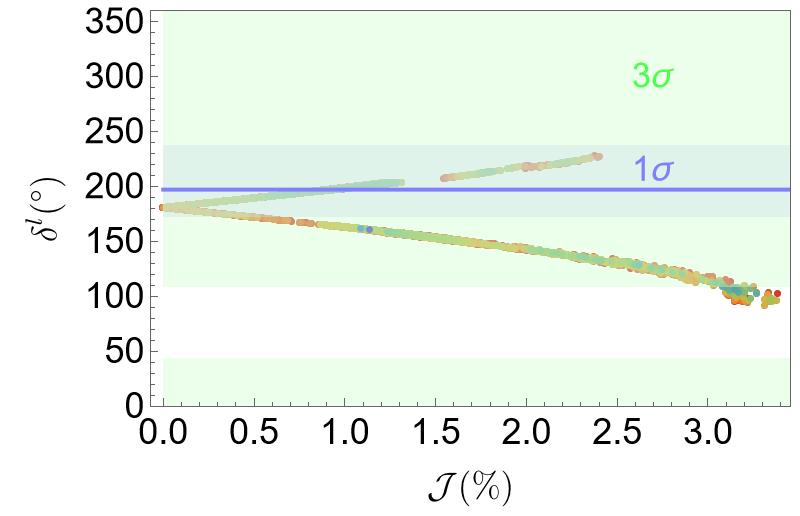}
		\includegraphics[width=0.45\textwidth]{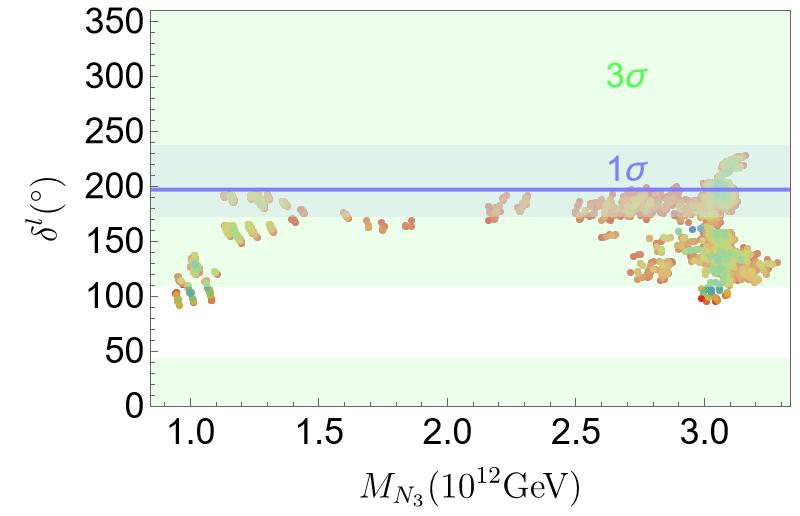}
		\includegraphics[width=0.4\textwidth]{fig/legend.pdf}
		\caption{The effective neutrino mass prediction (top left panel) and two-dimensional correlations between predicted observables for $\chi^2 \leq 10$.}
		\label{fig:prediction}
	\end{center}
\end{figure}

We perform a simple $\chi^2$ analysis with the $\chi^2$ function defined as 
\begin{eqnarray}
	\chi^2 = \sum_n \Big( \frac{{\rm obs}_n^{\rm th} ({\rm para}_m) - {\rm obs}_n^{\rm bf}}{\sigma({\rm obs}_n)} \Big)^2 \,,
\end{eqnarray} 
where ${\rm obs}_n$ run for all 9 independent observables in Eq.~\eqref{eq:predictions} and ${\rm para}_m$ account for all 8 free parameters in Eq.~\eqref{eq:parameters} with scanning regimes in Eqs.~\eqref{eq:scan_quark} and \eqref{eq:scan_lepton}.
We determine which regions fit the experimental data by setting an upper bound of $\chi^2$ value. The procedure of our scan is divided into two steps: First, we scan the quark sector by setting $\chi_q^2<10$. Then, we use the data of quark to run a subsequent scan and find points with $\chi_l^2<10$ in the charged lepton and neutrino sector. What's more, we further request that the $\chi^2 = \chi^2_q + \chi^2_l \leq 10$ and exclude points of $\chi^2 > 10$. The results are
shown in Figs.~\ref{fig:parameters} and ~\ref{fig:prediction}.

Fig.~\ref{fig:parameters} shows the dependence of $\chi^2$ value on the model parameters, as well as correlations among these parameters. We find that $c_\nu$ is restricted in the range (0,1.7) and $\chi^2$ reaches its minimal value 1.6 when the value of $c_\nu$ is around 0.1.\footnote{The $\chi^2$ value of this benchmark at NuFIT6.0 is 2.7.} The benchmark point at the minimal value of $\chi^2$ is listed in Appendix~\ref{app:2}.

We also observe a linear correlation between $r_1$ and $r_2$. This relationship can be analytically derived with the help of Eq.~\eqref{eq:Y_e} and Eq.~\eqref{eq:ABC_e},  
\begin{eqnarray}
	r_2 = \frac{4Y}{X}r_1+\frac{Z}{X} \,,
\end{eqnarray}
where $X=\zeta_e[-A_e+y_\tau+\eta_e(y_\mu-y_e)]+\zeta_d[A_d-y_b+\eta_d(y_d-y_s)], \, Y=A_u-y_t+\eta_u(y_u-y_c), \, Z=\zeta_e[-A_e+y_\tau+\eta_e(y_\mu-y_e)]+3\zeta_d[-A_d+y_b+\eta_d(y_s-y_d)]$. 
According to the rest panels in Fig.~\ref{fig:parameters}, the theory inputs have the following hierarchical relation:
\begin{eqnarray}
	A_d \ll A_u, \, m_R \ll m_L, \, r_1 \ll r_2 \,.
\end{eqnarray}

Fig.~\ref{fig:prediction} displays correlations between predicted observables. In the top left panel of Fig.~\ref{fig:prediction}, we display predictions for the effective mass $m_{ee}$ of the neutrinoless double beta decay vs the lightest neutrino $m_1$ for NO. The effective mass $m_{ee}$ is defined as:
\begin{eqnarray}
	m_{ee} = \left|\sum_{i=1}^{3} m_i (U_{\rm PMNS})_{ei}^2 \right| \,.
\end{eqnarray}
The predicted values of $m_{ee}$ are mainly distruibuted between 3.45 meV and 5.2 meV, and values of $m_1$ are smaller than 2.33 meV. 
Furthermore, we also calculate the CP-violating phase $\delta_l$ and the Jarlskog invariant $\mathcal{J}$, which measures the strength of CP violation in neutrino oscillations \cite{Jarlskog:1985ht,Fritzsch:1999im}. As illustrated in Fig.~\ref{fig:prediction}, {the CP-violating $\delta^l$ ranges in $(90^\circ,230^\circ)$}. $\mathcal{J}$ ranges form $0\%$ to $3.5\%$ and gets obviously larger as Dirac CP-violating phase $\delta^l$ deviates from $180^\circ$. At the benchmark point, $\delta^l$ is predicted to be $160^\circ$ and $\mathcal{J}$ is equal to $1.14\%$. The upcoming long-base neutrino experiments DUNE \cite{DUNE:2015lol} and Hyper-Kamiokande \cite{Hyper-Kamiokande:2018ofw} are supposed to give a resolution of $\delta_l$ at the level of $10^\circ$. Thus, they will be able to exclude a large parameter space of this model.

Once all free parameters are determined, we can get RH neutrino mass matrix through Eqs.~\eqref{eq:N_mass} and ~\eqref{eq:F}, of which eigenvalues are three RH neutrinos mass $M_{N_1}$, $M_{N_2}$ and $M_{N_3}$. As shown in Fig.~\ref{fig:prediction}, the heaviest RH neutrino mass is predicted to be around $(1\sim 3.3) \times 10^{12}$~GeV. We check in section \ref{sec:4} whether this is consistent with gauge unification and proton decay constraints.

\section{Gauge unification and proton decay}\label{sec:4}

$SO(10)$ GUTs have many breaking chains. In the context of a specific breaking chain, the solutions to the RGEs and the requirement of gauge unification impose restrictions on the GUT scale and establish a correlation with the intermediate scales. Since the proton decay rate depends on the GUT scale, we can utilize the limits on this observable to constrain both the GUT scale and intermediate scales. 
We recall our breaking chain is $
SO(10) \to G_{422}^{\rm C} \to G_{3221} \to G_{\rm SM}$, 
 as introduced in the beginning of section~\ref{sec:2}. 
In this analysis, we will go beyond the hypothesis of minimal particle contents and consider the influence of more Higgses on the gauge unification. 
On the other hand, the scale of the lowest intermediate symmetry breaking, denoted as $M_1$ in Eq.~\eqref{eq:breaking_chain}, refers to the $B-L$ breaking scale. Majorana masses for right-handed neutrinos are generated after the $B-L$  spontaneous breaking at $M_1$. By keeping the Yukawa couplings stay in the perturbative region, all right-handed neutrinos, including the heaviest one, should have masses not heavier than $M_1$. Below, we will check if the gauge unification in this model satisfies the experimental constraint from proton decay and see if there is a tension between the lowest intermediate scale and the predicted right-handed neutrino mass.

\subsection{Unification of gauge couplings}

Any intermediate symmetry between $SO(10)$ and SM can be expressed as a product of Lie groups $H_1\times\dots\times H_n$, the two-loop renormalisation group running equation for group $H_i$ is given by
\begin{eqnarray}
	\frac{{\rm d}\alpha_i(t)}{{\rm d}t}=\beta_i(\alpha_j)\,,
\end{eqnarray}
where $t=\log(\mu/\mu_0)$, $\alpha_i=g_i^2/4\pi$. $\beta$ function depends on the field contents of the theory:
\begin{eqnarray}
	\beta_i=\frac{1}{2\pi}\alpha_i^2(b_i+\frac{1}{4\pi}\sum_jb_{ij}\alpha_j)\,.
\end{eqnarray}
Here, $b_i$ and $b_{ij}$ are coefficients and can be written as
\begin{eqnarray}
	b_i &=& - \frac{11}{3} C_2(H_i) + \frac{2}{3} \sum_F T(\psi_i) + \frac{1}{3} \sum_S T(\phi_i) \,, \nonumber\\
	b_{ij} &=&
	- \frac{34}{3} [C_2(H_i)]^2 \delta_{ij} + \sum_F T(\psi_i) [2 C_2(\psi_j) + \frac{10}{3} C_2(H_i) \delta_{ij}] \nonumber\\
	&&+ \sum_S T(\phi_i) [4 C_2(\phi_j) + \frac{2}{3} C_2(H_i) \delta_{ij}]\,,
\end{eqnarray}
where the $\psi$ and $\phi$ indices sum over the fermions and complex scalar multiplets, respectively, and $\psi_i$ and $\phi_i$ are their representations in the group $H_i$, respectively. $C_2(R_i)$ (for $R_i=\psi_i, \varphi_i$) represents the quadratic Casimir of the representation $R_i$ in group $H_i$ and $C_2(H_i)$ is the quadratic Casimir of the adjoint presentation of the group $H_i$. If the condition $\frac{b_j}{2\pi}\alpha_j(t_0)(t-t_0)<1$ is satisfied, these equations can be analytically solved:
\begin{eqnarray}
	\alpha_i^{-1}(t)=\alpha_i^{-1}(t_0)-\frac{b_i}{2\pi}(t-t_0)+\sum_j\frac{b_{ij}}{4\pi b_i}\log\left(1-\frac{b_j}{2\pi}\alpha_j(t_0)(t-t_0)\right)\,.
\end{eqnarray}

The one-loop matching condition for a group $H_{i+1}$ broken to a subgroup $H_i$ at scale $M_I$ is given by \cite{Chakrabortty:2017mgi} $\alpha_{H_{i+1}}^{-1}(M_I)-\frac{1}{12\pi}C_2(H_{i+1}) = \alpha_{H_i}^{-1}(M_I)-\frac{1}{12\pi}C_2(H_i)$. When $SU(2)_R\times U(1)_X\longrightarrow U(1)_Y$, we have the matching condition \cite{Chakrabortty:2019fov} $\frac{3}{5}\Big(\alpha_{2R}^{-1}(M_I)-\frac{1}{6\pi}\Big)+\frac{2}{5}\alpha_{1X}^{-1}(M_I)=\alpha_{1Y}^{-1}(M_1)$. In this breaking pattern, ${\bf 54}_H,{\bf 45}_H,{\bf 10}_H,\overline{\bf 126}_H,{\bf 120}_H$ are needed to trigger symmetry breaking and generate fermion masses. In the following part, $n_H$ is used to represent the repetition number of the Higgs field ${\bf 10}_H,{\bf \overline{126}}_H,{\bf 120}_H$, the copy of ${\bf 54}_H,{\bf 45}_H$ is always set to one. For $M_2\to M_{\rm GUT}$, we assume only $({\bf 15,1,1})\subset {\bf 45}_H$ contributes to RG running, as it is responsible for the symmetry breaking from $SU(4)_c \times SU(2)_L \times SU(2)_R \times Z_2^{\rm C}$ to $SU(3)_c \times SU(2)_L \times SU(2)_R \times U(1)_X$, other components of ${\bf 45}_H$ are assumed to be around GUT scale. Therefore, when $n_H=0$, we have the following $\beta$-coefficients:
\begin{equation}
	\begin{aligned}
		&G_{422}^{\rm C}: \qquad \{b_i^0\} =  \left(
		\begin{array}{c}-\frac{28}{3} \\ -\frac{10}{3} \\ -\frac{10}{3}\\
		\end{array}
		\right)\,,\quad
		\{b_{ij}^0\} = \left(
		\begin{array}{ccc}
			-\frac{25}{6} & \frac{9}{2} & \frac{9}{2} \\
			\frac{45}{2} & \frac{11}{3} & 0 \\
			\frac{45}{2} & 0 & \frac{11}{3} \\
		\end{array}
		\right)\,,\\
		&G_{3221}: \qquad \{b_i^0\} =  \left(
		\begin{array}{c}-7 \\ -\frac{10}{3} \\ -\frac{10}{3} \\ 4 \\
		\end{array}
		\right)\,,\quad 
		\{b_{ij}^0\} = \left(
		\begin{array}{cccc}
			-26 & \frac{9}{2} & \frac{9}{2} & \frac{1}{2} \\
			12 & \frac{11}{3} & 0 & \frac{3}{2} \\
			12 & 0 & \frac{11}{3} & \frac{3}{2} \\
			4 & \frac{9}{2} & \frac{9}{2} & \frac{7}{2}
		\end{array}
		\right)\,.
	\end{aligned}
\end{equation}
The final $\beta$-coefficients can be obtained from $b_i=b_i^0+n_H\delta b_i,b_{ij}=b_{ij}^0+n_H\delta b_{ij}$. Next we discuss three scenarios where unification constraints are quite different.
\begin{itemize}
	\item[S1)] We assume the components of the Higgs multiplets that are unnecessary for symmetry breaking at lower scales to be heavy and decouple at higher scales. Minimal Higgs contents for each intermediate symmetry breaking is considered. Specifically, Higgs contents include $({\bf 1,2,2})\subset {\bf 10}_H,({\bf 15,2,2})+({\bf \overline{10},3,1})+({\bf 10,1,3})\subset {\bf \overline{126}}_H,({\bf 1,2,2})+({\bf 15,2,2})\subset {\bf 120}_H$ in $G_{422}^{\rm C}$, $({\bf 1,2,2},0)\subset {\bf 10}_H,({\bf 1,2,2},0)+({\bf 1,1,3},-1)\subset {\bf \overline{126}}_H,2({\bf 1,2,2},0)\subset {\bf 120}_H$ in $G_{3221}$. What's more, we assume Higgs bi-doublets $({\bf 1,2,2})\subset {\bf 10}_H$ and $({\bf 1,2,2})\subset {\bf 120}_H$ in $G_{422}^{\rm C}$ are mixing, $({\bf 1,2,2},0)\subset {\bf 10}_H$, $({\bf 1,2,2},0)\subset{\bf \overline{126}}_H$ and $2({\bf 1,2,2},0)\subset {\bf 120}_H$ in $G_{3221}$ are mixing too, therefore, there is only one $({\bf 1,2,2})$ and one $({\bf 1,2,2},0)$ contribute to gauge running. $\beta$-cofficients of these Higgs fields are
	\begin{equation}
		\begin{aligned}
			&G_{422}^{\rm C}: \qquad \{\delta b_i\} = \left(
			\begin{array}{c}\frac{50}{3} \\ 17 \\ 17\\
			\end{array}
			\right)\,, \quad
			\{\delta b_{ij}\} = \left(
			\begin{array}{ccc}
				\frac{2908}{3} & 168 & 168 \\
				840 & 321 & 93 \\
				840 & 93 & 321 \\
			\end{array}
			\right) \,,\\
			&G_{3221}: \qquad \{\delta b_i\} = \left(
			\begin{array}{c}0 \\ \frac{1}{3} \\ 1\\ \frac{3}{2}\\
			\end{array}
			\right)\,, \quad
			\{\delta b_{ij}\} = \left(
			\begin{array}{cccc}
				0 & 0 & 0 & 0 \\
				0 & \frac{13}{3} & 3 & 0 \\
				0 & 3 & 23 & 12 \\
				0 & 0 & 36 & 27
			\end{array}
			\right) \,.
		\end{aligned}
	\end{equation}
	\item[S2)] Except for the minimal Higgs contents in S1, we additionally add $({\bf 6,1,1})\subset {\bf 10}_H$ in $G_{422}^{\rm C}$ and its decomposition $({\bf 3,1,1},-\frac{1}{3}),({\bf \overline{3},1,1},\frac{1}{3})\subset {\bf 10}_H$ in $G_{3221}$. Higgs bi-doublets mixing is also considered here. $\beta$-cofficients of these Higgs fields are
	\begin{equation}
		\begin{aligned}
			&G_{422}^{\rm C}: \qquad \{\delta b_i\} = \left(
			\begin{array}{c}17 \\ 17 \\ 17\\
			\end{array}
			\right)\,, \quad
			\{\delta b_{ij}\} = \left(
			\begin{array}{ccc}
				982 & 168 & 168 \\
				840 & 321 & 93 \\
				840 & 93 & 321 \\
			\end{array}
			\right) \,,\\
			&G_{3221}: \qquad \{\delta b_i\} = \left(
			\begin{array}{c}\frac{1}{3} \\ \frac{1}{3} \\ 1\\ \frac{11}{6}\\
			\end{array}
			\right)\,, \quad
			\{\delta b_{ij}\} = \left(
			\begin{array}{cccc}
				\frac{22}{3} & 0 & 0 & \frac{2}{3} \\
				0 & \frac{13}{3} & 3 & 0 \\
				0 & 3 & 23 & 12 \\
				\frac{16}{3} & 0 & 36 & \frac{83}{3}
			\end{array}
			\right) \,.
		\end{aligned}
	\end{equation}
	\item[S3)] We assume all components of the Higgs multiplets in ${\bf 10}_H,\overline{\bf 126}_H,{\bf 120}_H$ contribute to the running of the gauge coupling. Higgs bi-doublets mixing is also considered here. $\beta$-cofficients of these Higgs fields are
	\begin{equation}
		\begin{aligned}
			&G_{422}^{\rm C}: \qquad \{\delta b_i\} = \left(
			\begin{array}{c}\frac{64}{3} \\ 21 \\ 21\\
			\end{array}
			\right)\,, \quad
			\{\delta b_{ij}\} = \left(
			\begin{array}{ccc}
				\frac{3584}{3} & 192 & 192 \\
				960 & 433 & 93 \\
				960 & 93 & 433 \\
			\end{array}
			\right) \,,\\
			&G_{3221}: \qquad \{\delta b_i\} = \left(
			\begin{array}{c}\frac{64}{3} \\ \frac{61}{3} \\ \frac{61}{3}\\ \frac{64}{3}\\
			\end{array}
			\right)\,, \quad
			\{\delta b_{ij}\} = \left(
			\begin{array}{cccc}
				\frac{2368}{3} & 192 & 192 & \frac{128}{3} \\
				512 & \frac{1273}{3} & 87 & 64 \\
				512 & 87 & \frac{1273}{3} & 64 \\
				\frac{1024}{3} & 192 & 192 & \frac{512}{3}
			\end{array}
			\right) \,.
		\end{aligned}
	\end{equation} 
\end{itemize} 

We scan the parameter space of $M_{\rm GUT},M_2,M_1$ allowed by gauge unification and calculate the gauge coupling $\alpha_{\rm GUT}$ at unification scale for each scenario. Results are shown in Fig~\ref{fig:RGE}. Obviously, as $n_H$ increases or equivalently adding more Higgs fields, $\beta$-cofficients become larger, which results in the gauge coupling $\alpha_{\rm GUT}$ evolves to a relatively large number, even becomes a landau pole. Comparing scenario S1 with scenario S2, we find $({\bf 6,1,1})\subset {\bf 10}_H$ in $G_{422}^{\rm C}$ and its decomposition $({\bf 3,1,1},-\frac{1}{3}),({\bf \overline{3},1,1},\frac{1}{3})\subset {\bf 10}_H$ in $G_{3221}$ can improve the GUT scale, but lower the lowest intermediate scale. Furthermore, when $n_H=3$, $\alpha_{\rm GUT}$ in scenario S2 grows faster than other two scenarios, so we demand $\alpha_{\rm GUT}\leq 0.4$ and cut off the part where $\alpha_{\rm GUT}\ge 0.4$, leading to the maximal value of $M_1$ gets small. As a result, in scenario S2, when $n_H=3$, $M_1$ is smaller than $M_{N_3}$ in Fig.~\ref{fig:prediction}. We also find $({\bf 1,2,2})$ and $({\bf 1,2,2},0)$ can improve the lowest intermediate scale but lower the GUT scale, so as to improve the GUT scale, we consider Higgs bi-doublets mixing and therefore reduce the contribution of Higgs bi-doublets to gauge running. In scenario S1 and S2, when $n_H=1$, there are not many Higgs fields, $\alpha_{\rm GUT}$ does not increase rapidly and ranges in (0.022,0.032). In scenario S3, there are already many Higgs fields when $n_H=1$, we find no solution which satisfy the energy scale hierarchy $M_1 \leq M_2 \leq M_{\rm GUT}$ when $n_H\ge 2$.

In conclusion, we should be careful about adding Higgs fields, more Higgs fields meaning faster evolution of gauge couplings and maybe evolve to a large number, which is not favored by theory. When $n_H=1$, the above three scenarios restrict $M_1$ should be smaller than $4.6\times 10^{13}$ GeV, consistent with $M_{N_3}<4.4\times 10^{13}$ GeV in Ref. \cite{King:2021gmj,Fu:2022lrn}. Furthermore, viable points of $\chi^2\leq10$ in our model always meet this requirement. 

\subsection{proton decay}\label{sec:4_2}

\begin{figure}[t!] 
	\begin{center} 
		\includegraphics[width=0.9\textwidth]{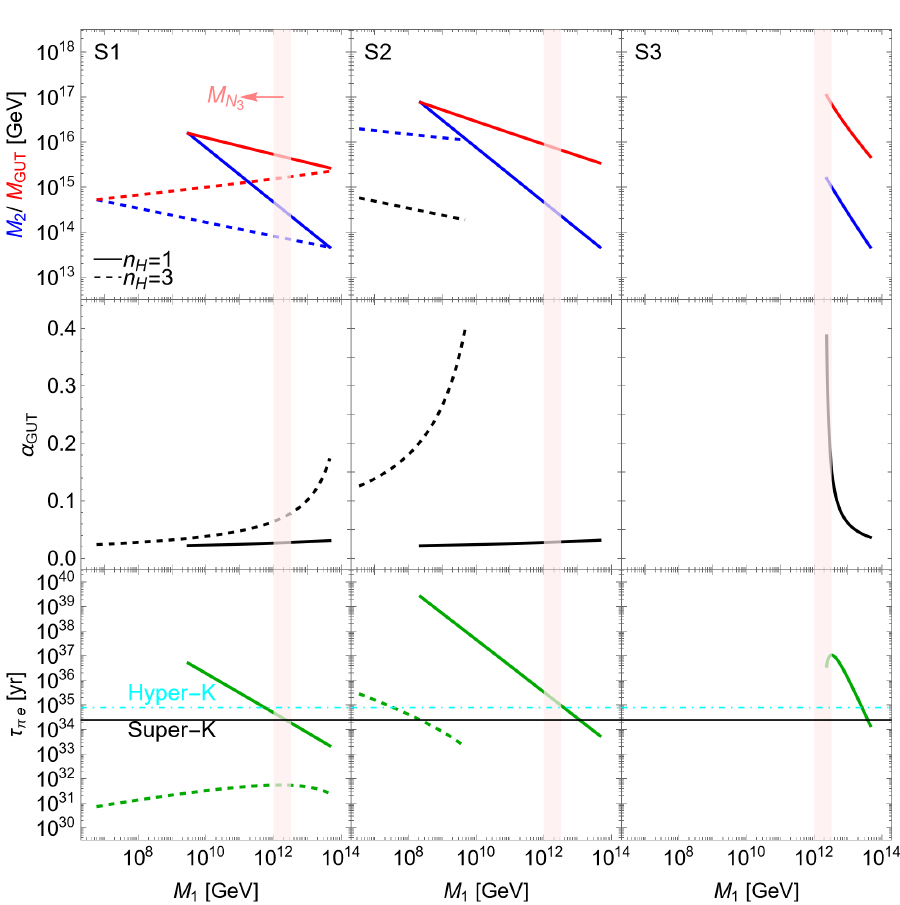}
		\caption{Predictions of GUT scale $M_{\rm GUT}$, intermediate scales, $M_1$ and $M_2$, gauge coupling at GUT scale $\alpha_{\rm GUT}$, and the partial lifetime for proton decaying to $\pi^0 e^+$ in three scenarios, S1 (left), S2 (middle) and S3 (right). The energy scale hierarchy $M_1 \leq M_2 \leq M_{\rm GUT}$ is required. $n_H = 1$ (solid curve) and $3$ (dashed curve) refer to the repetition number of the Higgs fields ${\bf 10}_H,{\bf \overline{126}}_H,{\bf 120}_H$. {The predicted range of $M_{N_3}$ in the last section is shown in the pink band as a reference}. S3 gives no solution for $n_H = 3$ and is thus not shown in the right panel.}
		\label{fig:RGE}
	\end{center}
\end{figure}

In non-SUSY $SO(10)$ GUTs, proton decay will be induced by integrating out the superheavy gauge fields $({\bf 3,2},-\frac{5}{6}),({\bf \overline{3},2},\frac{5}{6}),({\bf 3,2},\frac{1}{6}),({\bf \overline{3},2},-\frac{1}{6})$, which are typically denoted as $X^\mu,Y^\mu,X^{\prime\mu},Y^{\prime\mu}$, and resulting in the following four dimension-six operators \cite{FileviezPerez:2004hn}:
\begin{equation} \label{eq:D6}
	\begin{aligned}
		\epsilon^{ijk} \epsilon_{\alpha\beta} \Big(&\frac{1}{\Lambda_1^2}
		(\overline{u_R^{jc}} \gamma^\mu Q^k_\alpha)(\overline{d_R^{ic}} \gamma_\mu L_\beta) +
		\frac{1}{\Lambda_1^2}
		(\overline{u_R^{jc}} \gamma^\mu Q^k_\alpha)(\overline{e_R^{c}} \gamma_\mu Q_\beta^i)\\
		+ &\frac{1}{\Lambda_2^2}
		(\overline{d_R^{jc}} \gamma^\mu Q^k_\alpha)(\overline{u_R^{ic}} \gamma_\mu L_\beta) +
		\frac{1}{\Lambda_2^2}
		(\overline{d_R^{jc}} \gamma^\mu Q^k_\alpha)(\overline{\nu_R^{c}} \gamma_\mu Q^i_\beta)
		+{\rm h.c.} \Big) \,,
	\end{aligned}
\end{equation}
where $i,j,k$ ($\alpha,\beta$) denotes colour (flavour) indices and $\Lambda_{1}\simeq \sqrt{2}M_{(X,Y)}/g_{\rm GUT}, \Lambda_{2}\simeq \sqrt{2}M_{(X^\prime,Y^\prime)}/g_{\rm GUT}$ are the UV completion scales of the GUT symmetry. These four operators trigger proton decay in the form: $p\rightarrow M+\overline{l}$, where mesons $M$ can be $\pi^0,\pi^+,K^0,K^+,\nu$ and leptons $l$ can be $e,\mu,\nu$ \cite{Machacek:1979tx}. The partial decay width for such decay mode can be written as \cite{Aoki:2017puj,Saad:2022mzu}
\begin{equation} \label{eq:decay}
	\begin{aligned}
		\Gamma(p\rightarrow M+\overline{l})=\frac{m_p}{32\pi}\Big[1-\Big(\frac{m_M}{m_p}\Big)^2\Big]^2A_L^2\Big|\sum_{n}A_{Sn}W_nF_0^n(p\rightarrow M)\Big|^2 \,,
	\end{aligned}
\end{equation}
where $n=L,R$, $m_p$ and $m_M$ are masses of proton and Mesons, $W_n$ is the Wilson coefficients of the operators in Eq.~\eqref{eq:D6} that give rise to a specific decay channel, $F_0^n=\langle M | (qq^\prime)_{L,R} q^{\prime\prime}_L | p \rangle$ is the revelant hadronic matrix element and $q,q^\prime,q^{\prime\prime}=u,d,s$.

Furthermore, $A_L$ represents the long range effect from proton scale ($m_p\sim 1$ GeV) to electroweak scale ($M_Z$) which is calculated at two-loop level $A_L=1.247$ \cite{Nihei:1994tx,Ellis:2020qad}. $A_{SL}$ and $A_{SR}$ represent the short range effects obtained from RG running from $M_Z$ to the GUT scale $M_{\rm GUT}\simeq M_{(X,Y)}=M_{(X^\prime,Y^\prime)}$. Therefore, these two factors non-trivially denpend on the breaking chain. $A_{SL}$ and $A_{SR}$ are given by \cite{Buras:1977yy,Goldman:1980ah,Caswell:1982fx,Ibanez:1984ni}
\begin{eqnarray}
	A_{SL(R)} = \prod_A^{M_Z \leqslant M_A \leqslant M_X} \prod_i \left[ \frac{\alpha_i (M_{A+1})}{\alpha_i(M_A)} \right]^{\frac{\gamma_{iL(R)}}{b_i}}\,,
\end{eqnarray}
where $\gamma_i$ and $b_i$ denote the anomalous dimension and one-loop $\beta$ coefficient, respectively, and $\gamma_i$ at given intermediate scales can be found in \cite{King:2021gmj}.

We concentrate on the golden channel $p\rightarrow\pi^0e^+$, since $\Lambda_{1}\simeq\Lambda_{2}\simeq \sqrt{2}M_{\rm GUT}/g_{\rm GUT}$, decay widths of this channel can be written as
\begin{equation} \label{eq:pie}
	\begin{aligned}
		&\Gamma(p\rightarrow \pi^0 e_\alpha^+)=\frac{m_p}{32\pi}\Big[1-\Big(\frac{m_{\pi^0}}{m_p}\Big)^2\Big]^2A_L^2\frac{g_{\rm GUT}^4}{4M_{\rm GUT}^4}\times\\
		&\Big\{A_{SL}^2\Big|(U_u^{\prime T}U_u)_{11}(U_d^TU_e^\prime)_{1\alpha}+(U_u^{\prime T}U_d)_{11}(U_u^TU_e^\prime)_{1\alpha}\Big|^2\Big|\langle\pi^0|(ud)_{L}u_L|p\rangle\Big|^2+\\
		&A_{SR}^2\Big|(U_u^{\prime T}U_u)_{11}(U_d^{\prime T}U_e)_{1\alpha}+(U_d^{\prime T}U_u)_{11}(U_u^{\prime T}U_e)_{1\alpha}\Big|^2\Big|\langle\pi^0|(ud)_{R}u_L|p\rangle\Big|^2\Big\} \,.
	\end{aligned}
\end{equation}
Scan on the fermion masses and mixing allows us to determine each unitary matrix in this formula for each point in Fig.~\ref{fig:parameters}. 
In particular, as Yukawa coupling matrices are hermitian in our model, we have $U_u^\prime=U_u=P_uO_u,U_d^\prime=U_d=P_dO_d,U_e^\prime=U_e$, and $U_e$ can be derived throgh diagonalisation $U_e^\dagger Y_e U_e = \text{diag}\{y_e,y_\mu,y_\tau\}$. Once we substitute the numerical results into the above equation, proton decay lifetime of $p\rightarrow\pi^0e^+$ is only proportional to $\frac{4M_{\rm GUT}^4}{g_{\rm GUT}^4}=\Big(\frac{M_{\rm GUT}^2}{2\pi\alpha_{\rm GUT}}\Big)^2$. Predictions for the proton decay lifetime in this channel of the above three scenarios are displayed in Fig.~\ref{fig:RGE}. Here, we observe that the effect of flavor part $U_u,U_d,U_e$ in Eq.~\eqref{eq:pie} is small, the difference between the maximum and minimum value of proton decay lifetime corresponding to the same $M_1$ value is within 5\%. As $n_H$ increases or equivalently adding more Higgss fields, proton decay lifetime will decreases due to the increase in $\alpha_{\rm GUT}$. In scenario 2, we add $({\bf 6,1,1})\subset {\bf 10}_H$ in $G_{422}^{\rm C}$ and its decomposition $({\bf 3,1,1},-\frac{1}{3}),({\bf \overline{3},1,1},\frac{1}{3})\subset {\bf 10}_H$ in $G_{3221}$ to improve the GUT scale, and its proton decay lifetime can exceed the future Hyper-Kamiokande (HK) targeted $\tau_{\pi^0e^+}>7.8\times 10^{34}$ years \cite{Hyper-Kamiokande:2018ofw} even when $n_H=3$. However, the maximal value of the lowest intermediate scale $M_1$ allowed by Super-Kamiokande (SK) bound decreases to $2.3\times 10^{8}$ GeV, which is smaller than $M_{N_3}$ in Fig.~\ref{fig:prediction}. {When $n_H=1$, all three scenarios satisfy the SK bound $\tau_{\pi^0e^+}>2.4\times 10^{34}$ years \cite{Super-Kamiokande:2020wjk}, the constraints on $M_1$ of three scenarios are S1: $M_1\leq 2.6 \times 10^{12}$ GeV, S2: $M_1\leq 1.2 \times 10^{13}$ GeV, S3: $M_1\leq 3.9 \times 10^{13}$ GeV. It is clear that the predicted range of $M_{N_3}$ in Fig.~\ref{fig:prediction} consistently satisfy this requirement.}

\section{Conclusion}\label{sec:5}

We have proposed the universal two-zero texture (UTZT), which means all $3\times 3$ fermion Yukawa/mass matrices take two-zero flavour textures, in the $SO(10)$ GUT framework to restrict the flavour space of quarks and leptons. 
With a concrete example, we show that the UTZT flavour structure can be realised in a $Z_6$ flavour symmetry. The quark and lepton mass matrices are all correlated with each other due to the grand unification. The light neutrino mass matrix, generated via type-I+II seesaw mechanism, is proven to still maintain the UTZT property. Together with the relation between Dirac Yukawa coupling matrices in Eq.~\eqref{eq:Yukawa}, we scan {seven} free parameters to fit nine observables (three mixing angles and one CP-violating phase in the quark sector, three mixing angles and two mass-squared differences in the lepton sector). The leptonic CP-violating phase is regarded as a prediction in the range $(90^\circ, 230^\circ)$. The upcoming long-baseline neutrino experiments will have the potential to exclude a part of the parameter space. Our scan results show that the right-handed neutrino spectrum can be strongly restricted as the model has to fit all flavour data, including fermion masses, CKM mixing and PMNS mixing. The heaviest right-handed neutrino mass is predicted to be less than $3.3 \times 10^{12}$ GeV, which is allowed by gauge unification and proton decay measurements in non-SUSY $SO(10)$ GUTs. The predicted region of $m_{ee}$ vs $m_1$ in Fig.~\ref{fig:prediction} for neutrinoless double beta decay experiments will allow us to test grand unification. 

We also perform the gauge unification for a specific breaking chain with two intermediate scales and scan the range of the allowed intermediate scales. The colour triplet Higgses of $SU(3)_c$, which are decomposed from  sextet of $SU(4)_c$ and further decomposed from ${\bf 10}_H$ of $SO(10)$, are found to improve the GUT scale but lower the lowest intermediate scale. The bi-doublet Higgs of $SU(2)_L \times SU(2)_R$ can improve the lowest intermediate scale but lower the GUT scale. Therefore, we assume few copies of Higgs bi-doublets involving at low energy, leading to the improvement of GUT scale. When $n_H=1$ or equivalently not adding too many Higgs fields, all three scenarios above require that the maximal value of the lowest intermediate scale $M_1$ be less than $4.6\times 10^{13}$ GeV, which is consistent with predicted observables of this model in Fig.~\ref{fig:prediction}. Adding too many Higgs fields causes gauge coupling at GUT scale to become too large, which should be careful. 
Proton decay is also discussed. As long as Higgs fields are not added too many, gauge coupling at the GUT scale will not get very large, i.e., $\alpha_{\rm GUT}\in(0.022,0.032)$. Therefore, GUT scale is always high enough to meet the SK bound, i.e., roughly $M_{\rm GUT}\ge 4.5\times 10^{15}$ GeV.

\section*{Acknowledgement}

The authors would like to thank Z.-z. Xing and D. Zhang for useful discussions in the initial stage of this work. The authors are supported by National Natural Science Foundation of China (NSFC) under Grants Nos. 12205064, 12347103 and Zhejiang Provincial Natural Science Foundation of China under Grant No. LDQ24A050002.

\appendix

\section{General form of Yukawa matrices in $SO(10)$} \label{app:1}

We check the validity of Eq.~\eqref{eq:Yukawa} when ${\bf 10}_H$ and ${\bf 120}_H$ are complex, and the copy of ${\bf 10}_H$, $\overline{\bf 126}_H$, ${\bf 120}_H$ is $n > 1$. We start with the most general form of Yukawa terms in $SO(10)$, 
\begin{equation} \label{eq:Yukawa_couplings_complex}
	\begin{aligned}
		-{\cal L}_{\rm Y} &= (A_k)_{\alpha\beta} \; {\bf 16}^\alpha_F {\bf 16}^\beta_F {\bf 10}_H^k \; + \; (A'_k)_{\alpha\beta} \; {\bf 16}^\alpha_F {\bf 16}^\beta_F {\bf 10}_H^{k*} \; + \; 
		(B_k)_{\alpha\beta} \; {\bf 16}^\alpha_F {\bf 16}^\beta_F \overline{\bf 126}_H^k \;\\& + \; 
		i\, (C_k)_{\alpha\beta} \; {\bf 16}^\alpha_F {\bf 16}^\beta_F {\bf 120}_H^k \; + \; i\, (C'_k)_{\alpha\beta} \; {\bf 16}^\alpha_F {\bf 16}^\beta_F {\bf 120}_H^{k*} \; + \;{\rm h.c.}\,,
	\end{aligned}
\end{equation}
where $k=1,2,..., n$. The $A'_k$ and $C'_k$ terms are forbidden if the Peccei-Quinn (PQ) symmetry is imposed. ${\bf 10}_H$, ${\bf 120}_H$, $\overline{\bf 126}_H$ are decomposed to a series of electroweak doublets in the SM gauge symmetry and these doublets mix together. Up to now, we have totally $8n$ pairs of Higgs doublets
	\begin{equation}
		\begin{aligned}
			h_u=(10_{Hk}^u,120_{Hk}^{u1},120_{Hk}^{u2},\overline{126}_{Hk}^u)\,,\\ h_d=(10_{Hk}^d,120_{Hk}^{d1},120_{Hk}^{d2},\overline{126}_{Hk}^d)\,,
		\end{aligned}
	\end{equation}
	where superscripts $1$ and $2$ of $120_H^{u}$ and $120_H^{d}$ stand for SU(4) singlet and adjoint representation under the $G_{422}=SU(4)_c\times SU(2)_L\times SU(2)_R$ decomposition. 
	The Yukawa terms after decomposition are explicitly written as
		\begin{equation} \label{eq:fermion_masses_1}
		\begin{aligned}
			-{\cal L}_{\rm Y} &= (10_{Hk}^u A_k+ 10_{Hk}^{u*} A'_k)(qu^c+l\nu^c)+(10_{Hk}^d A_k+10_{Hk}^{d*} A'_k)(qd^c+le^c)\\&+\frac{1}{\sqrt{3}}\overline{126}_{Hk}^u B_k(qu^c-3l\nu^c)+\frac{1}{\sqrt{3}}\overline{126}_{Hk}^d B_k(qd^c-3le^c)\\&+(120_{Hk}^{u1} C_k+120_{Hk}^{u1*} C'_k)(qu^c+l\nu^c)+(120_{Hk}^{d1} C_k+120_{Hk}^{d1*} C'_k)(qd^c+le^c)\\&-\frac{1}{\sqrt{3}}(120_{Hk}^{u1} C_k+120_{Hk}^{u1*} C'_k)(qu^c-3l\nu^c)+\frac{1}{\sqrt{3}}(120_{Hk}^{d2} C_k+120_{Hk}^{d2*} C'_k)(qd^c-3le^c)\,.
		\end{aligned}
	\end{equation}
	
	It is convenient to rotate the interaction basis $h$ to the mass basis $\hat{h}$ via a unitary transformation, 
	\begin{eqnarray} \label{eq:mix}
	h_a =  \begin{pmatrix} \widetilde{h_u} \\ h_d \end{pmatrix}_a \to \hat{h}_i =  W_{i,a} h_a \,,
	\end{eqnarray}
	where, $\widetilde{h_u} = i\sigma_2 h_u^*$,  $W$ is a unitary matrix,  the subscript for mass states $i=1,2,..., 8n$ is arranged following the mass ordering, and the subscript for interaction states $a = j_k$ is further split into two subscript $j =1,2,...8$ and $k = 1,2, ..., n$ with respect to the copy $n>1$. 
	 In the minimal case, only the SM higgs $h_{\rm SM} \equiv \hat{h}_1$, which is the massless Higgs doublet before electroweak symmetry breaking, contributes to fermion masses. 
	Then, we decompose Yukawa couplings in Eq.~\eqref{eq:fermion_masses_1} into its SM parts and get terms of fermion masses
	\begin{equation} \label{eq:fermion_masses}
		\begin{aligned}
			-{\cal L}_{\rm Y} &=
			(W_{1,1_k} A_k+W^*_{1,1_k} A'_k)h_{\rm SM}(qu^c+l\nu^c)+(W^*_{1,5_k} A_k+W_{1,5_k} A'_k)h_{\rm SM}(qd^c+le^c)\\&+\frac{1}{\sqrt{3}}W_{1,4_k} B_k h_{\rm SM}(qu^c-3l\nu^c)+\frac{1}{\sqrt{3}}W^*_{1,8_k} B_k h_{\rm SM} (qd^c-3le^c) \\&+(W_{1,2_k} C_k+W^*_{1,2_k} C'_k)h_{\rm SM}(qu^c+l\nu^c)+(W^*_{1,6_k} C_k+W_{1,6_k} C'_k)h_{\rm SM}(qd^c+le^c)\\&-\frac{1}{\sqrt{3}}(W_{1,3_k} C_k+W^*_{1,3_k} C'_k) h_{\rm SM}(qu^c-3l\nu^c)\\&+\frac{1}{\sqrt{3}}(W^*_{1,7_k} C_k+W_{1,7_k} C'_k)h_{\rm SM}(qd^c-3le^c)\,.
		\end{aligned}
	\end{equation}
	Based on these terms, we can parameterize the Dirac Yukawa coupling matrices of fermions into the following more concise form
	\begin{equation}
		\begin{aligned}
			Y_u&=H+H'+r_2 F+i(r_3 G+r^*_3 G')\,,\\
			Y_d&=r_1 H+r^*_1 H'+ r_1F+i(r_1 G+r^*_1G')\,,\\
			Y_\nu&=H+H'-3r_2 F+i(c_\nu G+c^*_\nu G')\,,\\
			Y_e&=r_1 H+r^*_1 H'-3r_1F+i(c_e r_1 G+c_e^* r^*_1 G')\,,
		\end{aligned}
		\label{eq:Yukawa_matrices_parameterize}
	\end{equation}
	where
	\begin{eqnarray}
		\begin{aligned}
			&H=W_{1,1_k} A^*_k,\ H'=W^*_{1,1_k} A^{\prime *}_k,\ F=\frac{W^*_{1,8_k}}{\sqrt{3}r_1} B^*_k,\\&G=-i\frac{W^*_{1,6_k} +\frac{1}{\sqrt{3}}W^*_{1,7_k}}{r_1}C^*_k,\ G'=-i\frac{W_{1,6_k} +\frac{1}{\sqrt{3}}W_{1,7_k}}{r^*_1}C^{\prime *}_k,\\
			&r_1=\frac{W^*_{1,5_k}}{W_{1,1_k}},\  r_2=\frac{W_{1,4_k}}{W^*_{1,8_k}}r_1, \ r_3=\frac{W_{1,2_k}-\frac{1}{\sqrt{3}}W_{1,3_k}}{W^*_{1,6_k}+\frac{1}{\sqrt{3}}W^*_{1,7_k}}r_1,\\
			&c_e=\frac{W^*_{1,6_k}-\frac{3}{\sqrt{3}}W^*_{1,7_k}}{W^*_{1,6_k}+\frac{1}{\sqrt{3}}W^*_{1,7_k}},\ c_\nu=\frac{W_{1,2_k}+\frac{3}{\sqrt{3}}W_{1,3_k}}{W^*_{1,6_k} +\frac{1}{\sqrt{3}}W^*_{1,7_k}}r_1\,.
		\end{aligned}
		\label{eq:HFG}
	\end{eqnarray}
	If the PQ symmetry is imposed, $A'_k$ and $C'_k$ are forbidden, and then we arrive at Eq~\eqref{eq:Yukawa}. On the other hand, if $A'_k$ and $C'_k$ are present and $r_1,r_3,c_e,c_\nu$ are real numbers, Eq.~\eqref{eq:Yukawa_matrices_parameterize} is simplified as 
	\begin{equation}
		\begin{aligned}
			Y_u&=(H+H')+r_2 F+ir_3(G+G')\,,\\
			Y_d&=r_1(H+H')+ r_1F+ir_1(G+G')\,,\\
			Y_\nu&=(H+H')-3r_2 F+ic_\nu(G+G')\,,\\
			Y_e&=r_1 (H+H')-3r_1F+ic_e r_1(G+G')\,,
		\end{aligned}
		\label{eq:Yukawa_matrices_parameterize_2}
	\end{equation}
	We arrive at the same form as Eq~\eqref{eq:Yukawa} again. In fact, when introduce ${\bf 10}_H$, $\overline{\bf 126}_H$, ${\bf 120}_H$ to generate fermion masses, as $(\bf 1,2,2)\in {\bf 10}_H,{\bf 120}_H$ and $(\bf 15,2,2)\in {\bf 120}_H,\overline{{\bf 126}}_H$ are responsible for electroweak symmetry breaking and give fermion masses, these four Higgs fields only lead to totally eight independent terms in Yukawa coupling matrices, four terms $W_{1,1_k} A^*_k+W^*_{1,1_k} A^{\prime *}_k,W_{1,4_k} B^*_k,W_{1,2_k} C^*_k+W^*_{1,2_k} C^{\prime *}_k,W_{1,3_k} C^*_k+W^*_{1,3_k} C^{\prime *}_k$ in $Y_u,Y_\nu$, four terms $W^*_{1,5_k} A^*_k+W_{1,5_k} A^{\prime *}_k,W^*_{1,8_k} B^*_k$, $W^*_{1,6_k} C^*_k+W_{1,6_k} C^{\prime *}_k,W^*_{1,7_k} C^*_k+W_{1,7_k} C^{\prime *}_k$ in $Y_d,Y_e$. 
	
	The discussion can also be extended to the case of more Higgses evolving in the fermion masses. For example, in the two-Higgs-doublet model (THDM),  by denoting the two light Higgses as $\hat{h}_1=W_{1,a}h_a,\hat{h}_2=W_{2,a} h_a$, the above eight terms will be rewritten by replacing $W^*_{1,j_k}$ with certain combinations of $W^*_{1,j_k}$ and $W^*_{2,j_k}$, which will not be repeated here. 
	
In the case of real ${\bf 10}_H$ and ${\bf 120}_H$, ${\bf 10}_H^*$ is identical to ${\bf 10}_H$, so is ${\bf 120}_H$. terms of $A',C'$ can be absorbed into terms of $A$ and $C$, respectively. Furthermore, since $\widetilde{10_{Hk}^u} = 10_{Hk}^d$, $\widetilde{120_{Hk}^{u1}} = 120_{Hk}^{d1}$, and $\widetilde{120_{Hk}^{u2}} = 120_{Hk}^{d2}$. The total copy of doublets is reduced to $5n$ and thus the subscript $a$ runs for $1_k, 2_k, ..., 5_k$.

\section{Benchmark study} \label{app:2}

Among all points in our scan, we find that the minimal value of $\chi^2$ is 1.6. Inputs and predictions of fermion Yukawa matrices and mixing parameters are shown in Table~\ref{tab:benchmark_chisq}. Yukawa matrices $H$, $F$ and $G$ at this point are given by
\begin{eqnarray}\label{eq:benchmark_Yukawa}
	H &=& 10^{-2} \cdot \left(
	\begin{array}{ccc}
		0 & -0.0745 & 0 \\
		-0.0745 & 6.18 & -13.13 \\
		0 & -13.13 & 33.4 \\
	\end{array}
	\right)\,, \nonumber\\
	F &=& 10^{-2} \cdot \left(
	\begin{array}{ccc}
		0 & -0.1104 & 0 \\
		-0.1104 & -0.1588 & 1.898 \\
		0 & 1.898 & -4.781 \\
	\end{array}
	\right)\,, \nonumber\\
	G &=& 10^{-2} \cdot \left(
	\begin{array}{ccc}
		0. & -0.0074 & 0 \\
		0.0074 & 0. & -4.74 \\
		0 & 4.74 & 0. 
	\end{array}
	\right)\,,
\end{eqnarray}
respectively. 
From this, all charged fermion Yukawa matrices and the light neutrino mass matrix are obtained:
\begin{eqnarray}
	Y_u &=& \left(
	\begin{array}{ccc}
		0 & (0.09-6.37i) \cdot 10^{-5} & 0 \\
		(0.09+6.37i) \cdot 10^{-5} & 0.0628 & -0.1441-0.0411i \\
		0 & -0.1441+0.0411i & 0.365 \\
	\end{array}
	\right)\,, \nonumber\\
	Y_d &=& 10^{-2} \cdot \left(
	\begin{array}{ccc}
		0 & 0.0031+0.0001i & 0 \\
		0.0031-0.0001i & -0.1010 & 0.1885+0.0796i \\
		0 & 0.1885-0.0796i & -0.481 \\
	\end{array}
	\right)\,, \nonumber\\
	Y_e &=& 10^{-2} \cdot \left(
	\begin{array}{ccc}
		0 & -0.0043+0.0003i & 0 \\
		-0.0043-0.0003i & -0.1167 & 0.316-0.212i \\
		0 & 0.316+0.212i & -0.802 \\
	\end{array}
	\right)\,, \nonumber\\
	Y_\nu &=& 10^{-2} \cdot \left(
	\begin{array}{ccc}
		0 & -0.298-0.001i & 0 \\
		-0.298+0.001i & 5.86 & -9.28-0.44i \\
		0 & -9.28+0.44i & 23.73 \\
	\end{array}
	\right)\,, \nonumber\\
	M_{\nu} &=& 10^{-2} \cdot \left(
	\begin{array}{ccc}
		0 & 1.074 & 0 \\
		1.074 & -3.72-1.50i & -0.246+1.909i \\
		0 & -0.246+1.909i & 1.844 \\
	\end{array}
	\right)~{\rm eV}\,.
\end{eqnarray}
Then, applying the inverse of the type-(I+II) seesaw formula, the right-handed neutrino mass matrix is
\begin{alignat}{2} \label{eq:right_neutrino}
	& M_{\nu_R} =10^{12} \cdot \left(
	\begin{array}{ccc}
		0 & 0.0598 & 0 \\
		0.0598 & 0.0860 & 1.028 \\
		0 & 1.028 & 2.589 \\
	\end{array}
	\right) ~{\rm GeV} \,.
\end{alignat}
Three RH neutrino masses are predicted of order $10^{10}$, $10^{11}$, $10^{12}$~GeV, respectively. In particular, heaviest right-handed neutrino mass is $2.96 \times 10^{12}$ GeV, which is below the $B-L$ breaking scale $M_{B-L} \simeq  4.6 \times 10^{13}$ GeV $(n_H=1)$ and thus consistent with proton decay measurements. 

\begin{table}[t!] 
	\centering
	\begin{tabular}{m{2cm}<{\centering}|m{2.5cm}<{\centering}|m{2.5cm}<{\centering}|m{2.5cm}<{\centering}|m{2.5cm}<{\centering}}
		\hline \hline
		\multirow{6}{*}{Inputs} & $A_u$ & $A_d$ & $\phi_1$ & $\phi_2$ \\
		& 0.3665 & 0.0048 & 4.7379 & 1.4232 \\
		\cline{2-5}
		& $c_\nu$ & $m_L$ & $m_R$ & \\
		& 0.0940 & 13.8057 eV & 0.5588 eV & \\
		\cline{2-5}
		& \multicolumn{4}{c}{$(\eta_u,\eta_d,\eta_e,\zeta_d)$} \\
		& \multicolumn{4}{c}{(+, +, \textendash, \textendash)} \\
		\hline
		\multirow{4}{*}{Outputs} & $\theta_{12}^q$ & $\theta_{23}^q$ & $\theta_{13}^q$ & $\delta^q$ \\
		& $13.0356^\circ$ & $2.8011^\circ$ & $0.2366^\circ$ & $67.04^\circ$ \\
		\cline{2-5}
		& $\theta_{12}^l$ & $\theta_{23}^l$ & $\theta_{13}^l$ & $\delta^l$ \\
		& $33.26^\circ$ & $48.52^\circ$ & $8.55^\circ$ & $160^\circ$ \\
		\cline{2-5}
		\multirow{4}{*}{$\chi^2=1.6$} & $m_1$ & $\Delta m_{21}^2$ & $\Delta m_{31}^2$ & $\left<m\right>_{ee}$ \\
		& 0.491 meV & $7.44 \cdot 10^{-5}$ & $2.509 \cdot 10^{-3}$ & 3.98 meV \\
		\cline{2-5}
		& $M_{N_1}$ & $M_{N_2}$ & $M_{N_3}$ & $\mathcal{J}$ \\
		& $1.07\!\cdot\! 10^{10} \, \rm GeV$ & $2.93\!\cdot\! 10^{11} \, \rm GeV$ & $2.96\!\cdot\! 10^{12} \, \rm GeV$ & 1.14\% \\
		\hline
		\hline
	\end{tabular}
	\caption{Inputs and predictions of neutrino masses and mixing parameters for minimal $\chi^2$ in our scan. Charged fermion masses are all fixed at experimental best-fit values. Neutrino masses with normal ordering are predicted.}
	\label{tab:benchmark_chisq}
\end{table}

\end{document}